\documentclass[structabstract]{aa}  

\usepackage{verbatim}
\usepackage{lscape}
\usepackage[british]{babel}
\usepackage{natbib}
\bibpunct{(}{)}{;}{a}{}{,} 

\usepackage{graphicx}
\usepackage{url}
\usepackage{txfonts}

\begin{document}

\title{$^7$Li surface abundance in pre-MS stars.}
\subtitle{Testing theory against clusters and binary systems.}

\author{
  E. Tognelli \inst{1}
  \and
  S. Degl'Innocenti \inst{1,2}
  \and
  P. G. Prada Moroni \inst{1,2}
  }

\institute{
  Physics Department ``E. Fermi'', University of Pisa, largo B. Pontecorvo 3, I-56127, Pisa, Italy\\ 
  \and
  INFN, largo B. Pontecorvo 3, I-56127, Pisa, Italy\\
  }

\offprints{Emanuele Tognelli, tognelli@df.unipi.it}
\date{Received 24 February  2012 / accepted 8 October 2012}

\abstract
   {The disagreement between theoretical predictions and observations for surface lithium abundance in stars is a long-standing problem, which indicates that the adopted physical treatment is still lacking in some points. However, thanks to the recent improvements in both models and observations, it is interesting to analyse the situation to evaluate present uncertainties.}
   {We present a consistent and quantitative analysis of the theoretical uncertainties affecting surface lithium abundance in the current generation of models.}
   {By means of an up-to-date and well tested evolutionary code, \texttt{FRANEC}, theoretical errors on surface $^7$Li abundance predictions, during the pre-main sequence (pre-MS) and main sequence (MS) phases, are discussed in detail. Then, the predicted surface $^7$Li abundance was tested against observational data for five open clusters, namely Ic 2602, $\alpha$ Per, Blanco1, Pleiades, and Ngc 2516, and for four detached double-lined eclipsing binary systems. Stellar models for the aforementioned clusters were computed by adopting suitable chemical composition, age, and mixing length parameter for MS stars determined from the analysis of the colour-magnitude diagram of each cluster. We restricted our analysis to young clusters, to avoid additional uncertainty sources such as diffusion and/or radiative levitation efficiency.}
   {We confirm the disagreement, within present uncertainties, between theoretical predictions and $^7$Li observations for standard models. However, we notice that a satisfactory agreement with observations for $^7$Li abundance in both young open clusters and binary systems can be achieved if a lower convection efficiency is adopted during the pre-MS phase with respect to the MS one.}
   {}
   

\maketitle

\section{Introduction}

In the last two decades, a large number of $^7$Li observations have been collected for isolated stars, binary systems, and open clusters from the pre-MS to the late MS phases \citep[see e.g. Table 1 and references therein in][]{jeffries00,sestito05}, showing that $^7$Li depletion is a strong function of both mass and age. A detailed and homogeneous analysis has been carried out by \citet{sestito05}, who determined surface $^7$Li abundance for a large sample of open clusters in a wide range of ages and chemical compositions, supplying a useful tool for accurately analysing the temporal evolution of surface $^7$Li abundance. 

Open clusters and detached double-lined eclipsing binaries (EBs) are ideal systems for testing the validity of stellar evolutionary models, since their members have the same chemical composition and age. As a consequence, they allow the different lithium depletion pattern to be investigated as a function of the stellar mass once the age and the chemical composition have been kept fixed. 

Besides the large amount of $^7$Li data available, a strong effort in theoretical modelling has been made in the past years, and many different theoretical scenarios have been proposed to explain the observed surface $^7$Li abundance and its temporal evolution \citep[see e.g. the reviews in][]{deliyannis00,pinsonneault00,charbonnel00}, both in the framework of \emph{standard} and \emph{non-standard models} \citep[see e.g.,][]{pinsonneault90,pinsonneault94,chaboyer95,dantona97,ventura98,piau02,dantona03,montalban06}. 

\emph{Standard models} assume a spherically symmetric structure and convection and diffusion are the only processes that mix surface elements with the interior. Although the validity of such models in reproducing the main evolutionary parameters has been largely tested against observations, they fail to reproduce the observed $^7$Li abundances. Indeed, standard models show a $^7$Li depletion during the pre-MS phase that is much stronger than observed, while the opposite occurs in the MS phase \citep[see e.g.,][]{jeffries00}. Moreover, they cannot fully account for the formation of the so-called lithium dip for MS stars in the temperature range $6000\,\mathrm{K}\la T_\mathrm{eff}\la 7000\,\mathrm{K}$ \citep{boesgaard86}, see e.g. \citet{richer93}. 

The comparison between theory and observation is improved, in some cases, by introducing \emph{non-standard} processes into the models, e.g. rotation, gravity waves, magnetic fields, and accretion/mass loss \citep{pinsonneault90,dantona93,chaboyer95,talon98,ventura98,mendes99,siess99,dantona00,charbonnel05,baraffe10,vick10}. All these processes produce structural changes, with a related strong effect on lithium abundance \citep[see e.g. the reviews by][]{charbonnel00,talon08,talon10}. In particular, models with rotation-induced mixing plus gravity waves are able to reproduce $^7$Li the depletion during the MS and post MS phases \citep[i.e. the lithium dip feature and red-giant branch abundances, see e.g.,][]{talon10,pace12}.

A crucial point in stellar modelling, both for standard and non-standard models, concerns the treatment of the over-adiabatic convection efficiency in the stellar envelope, which is an important issue for lithium depletion, too.  In evolutionary codes, the most widely used convection treatment is the simplified scheme of the \emph{mixing length theory} \cite[MLT,][]{bohm58}. In this formalism, convection efficiency depends on a free parameter to be calibrated. It is a common approach to calibrate it by reproducing the solar radius. This choice usually gives good agreement between models and photometric data; however, to reproduce the effective temperature of stars with different masses in different evolutionary phases, an ad hoc value of the mixing length parameter should be adopted, as suggested by observations \citep[see e.g.,][]{chieffi95,morel00,ferraro06,yildiz07,gennaro11,piau11,bonaca12} and detailed hydrodynamical simulations \citep[see e.g.,][]{ludwig99,trampedach07}.

The main goal of this paper is to re-examine the old lithium problem in light of the improvements in the adopted physical inputs and observational data and to perform a quantitative analysis of the uncertainties affecting surface lithium depletion during the pre-MS phase. The aim is to compute, by means of updated models, theoretical error bars to be applied to the comparison between predictions and data available for stars in young open cluster and binary systems, as partially done in earlier other works \citep[see e.g.,][]{dantona84,swenson94,ventura98,piau02,sestito06}.

The paper is structured in the following way. Section \ref{sec:data} presents the adopted  $^7$Li data sample for the selected open clusters, followed by a brief description of present models (Sect. \ref{sec:models}). In Sect. \ref{sec:error} we evaluate the main  theoretical uncertainties affecting surface lithium abundance. Finally, in Sect. \ref{sec:results}, the comparison between predicted and observed lithium abundances for both young open clusters and binary systems is discussed.

\section{Lithium data}
\label{sec:data}

Surface $^7$Li abundances for young open clusters are taken from the homogeneous database made available by \citet{sestito05}. Here, we focus our analysis on clusters younger than about 150 - 200 Myr, in order to avoid MS depletion effects \citep[see e.g.,][]{sestito05}, with different metallicities for which a significant number of data in a wide range of effective temperatures are available. The clusters that satisfy these criteria are, Ic 2602, $\alpha$ Per, Blanco 1, Pleiades, and Ngc 2516.

Lithium abundances for young double-lined eclipsing binaries are not present in the database by \citet{sestito05}, but they have been measured by different authors, as we discuss in Sect. \ref{sec:binary}.

\section{Theoretical stellar models}
\label{sec:models}

Present stellar models were computed with an updated version of the \texttt{FRANEC} evolutionary code \citep{deglinnocenti08}, which adopts the most recent input physics, as described in detail by \citet{tognelli11}. The initial deuterium mass fraction abundance is fixed to $X_{\mathrm{D}} = 2\times 10^{-5}$ as a representative value for population I stars \citep[see e.g.][]{geiss98,linsky06,steigman07}. The logarithmic initial lithium abundance is assumed to be $\epsilon_{\mathrm{Li}} = 3.2 \pm 0.2$ \citep[see e.g.,][]{jeffries06,lodders09}, which approximatively corresponds to $X_{^7\mathrm{Li}} \approx 7\times 10^{-9}$ - $1\times 10^{-8}$ in dependence on the metallicity adopted for the models\footnote{We adopt a simple scaling of the initial $^7$Li abundance with the metallicity because we are mainly interested in reproducing the lithium depletion pattern, i.e.  $\epsilon_{\mathrm{Li}} - T_\mathrm{eff}$, which is independent of the initial $^7$Li abundance.}.

 Convection is treated according to the mixing length theory, using the same formalism presented in \citet{cox}. The adopted reference value of mixing length parameter is $\alpha = 1.0$ (as suggested by present comparison with pre-MS data, see Sect. \ref{sec:ammassi}).

\section{Theoretical uncertainties}
\label{sec:error}

\subsection{Chemical composition}
\label{sec:err_chim}

\begin{figure*}[t]
\centering
\includegraphics[width=0.9\linewidth]{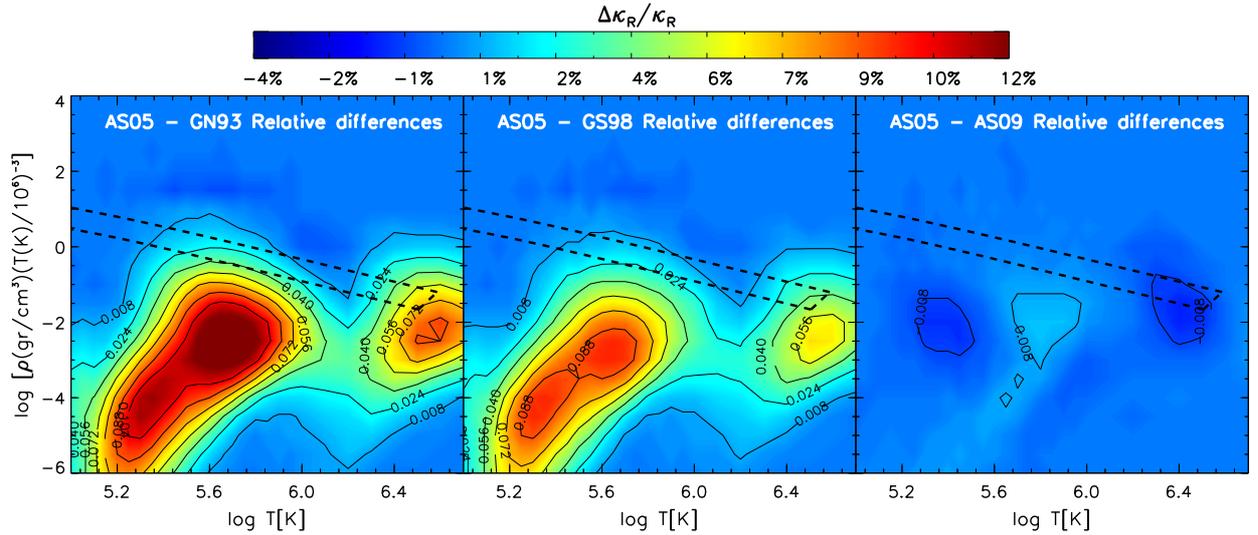}
\caption{Relative differences of the Rosseland radiative opacity coefficients ($\Delta \kappa_\mathrm{R}/\kappa_\mathrm{R}$) computed by adopting the AS05 and GN93 (\textit{left panel}), AS05 and GS98 (\textit{central panel}), and AS05 and AS09 (\textit{right panel}) solar mixtures, for Z =0.0129 and Y = 0.274. The three panels also show  the region covered by the entire convective envelope of masses in the range 0.6 - 1.2 M$_\odot$ (\textit{thick-dashed box}), for the same chemical composition of the opacity tables and adopting the mixing length value $\alpha = 1.00$.}
\label{fig:misture}
\end{figure*}

To properly calculate pre-MS evolution suitable initial abundances of helium, light elements, and metals are needed. For most of the stars, however, only the [Fe/H] value is available, so theoretical or semi-empirical assumptions are required.

Assuming for Population I stars a solar-scaled heavy elements distribution \citep[see e.g.,][for a detailed review]{asplund09}, the $Z/X$ value currently present at the stellar surface can be directly inferred from the observed [Fe/H]. For all the stars analysed in the paper, this value can be safely adopted as a good approximation  of the initial one over the whole structure, since the effect of microscopic diffusion is negligible owing to the very young ages involved. The initial helium content of the star $Y$ cannot be directly measured in the stellar spectra of cool stars, so a further relation between the initial metallicity and helium of the star is required. A common way to proceed is to assume the following linear relation \citep[see e.g.,][]{gennaro10}:  
\begin{equation}
Y = Y_{\mathrm{P}} + \frac{\Delta Y}{\Delta Z} Z
\label{eq:elio}
\end{equation}
where $Y_{\mathrm{P}}$ and $\Delta Y/\Delta Z$ represent, respectively, the primordial helium abundance and the helium-to-metal enrichment ratio. For the calculations we adopt $Y_\mathrm{p}=0.2485 \pm 0.0008$ \citep{cyburt04} and $\Delta Y/\Delta Z = 2\pm 1$ \citep{casagrande07}. Thus, the metallicity of the star can be obtained directly from the following equation,
\begin{equation}
Z = \frac{(1-Y_{\mathrm{P}})(Z/X)_\odot }{10^{-[\mathrm{Fe/H}]}-(1+\Delta Y/\Delta Z)(Z/X)_\odot}\nonumber\\
\label{eq:metal}
\end{equation} 
once the solar $(Z/X)_\odot$ has been specified. Regarding this last quantity, there are several values adopted by different authors, i.e. the still widely adopted \citet{grevesse93} (GN93, $(Z/X)_\odot = 0.0244$), \citet{grevesse98} (GS98,$(Z/X)_\odot~=~0.0231$) and the recent determinations by \citet{asplund05} (AS05, $(Z/X)_\odot = 0.0165$) and \citet{asplund09} (AS09, $(Z/X)_\odot =0.0181$), which are based on detailed 3D hydrodynamical atmosphere models. Recently, \citet{caffau10} (CL10) have found a value for the solar carbon photospheric abundance higher by about 0.1 dex than the previous one derived by AS09. This also leads to an increase in the solar metallicity-to-hydrogen ratio, namely $(Z/X)_\odot = 0.0211$, which is higher than the AS09 and much closer to the GS98 one.

Our models are computed adopting the AS05 mixture, for consistency with the extended pre-MS tracks and isochrones database already provided by our group\footnote{The database contains a very large grid of pre-MS models and isochrones between 1 - 100 Myr \citep[for more details see,][]{tognelli11}. The corresponding database is available at: \url{http://astro.df.unipi.it/stellar-models/}} \citep{tognelli11}, whereas for the conversion of [Fe/H] into ($Y$, $Z$) we prefer to use $(Z/X)_\odot=0.0181$ from the more recent AS09 heavy elements distribution. The inconsistency that may arise is negligible. Indeed, we verified that the effect on pre-MS models of adopting the AS05 or AS09 distribution in the opacity, once $Z$ and $Y$ have been kept fixed, is much lower than the variation produced by a change of ($Y$, $Z$) related to the error on $(Z/X)_\odot$ \citep[see][and the following discussion]{tognelli11}.

For the uncertainty on $(Z/X)_\odot$, the commonly suggested value is about $\pm 15\%$ \citep[see e.g.,][]{bahcall04,bahcall05}. However,  to take the difference between recent $(Z/X)_\odot$ determinations into account, i.e. GS98, AS09, and CL10, a larger uncertainty of about $+25\%$ with respect to AS09 is required. Thus, we use a final uncertainty of $+25$/$-15\%$ on $(Z/X)_\odot$.

Besides the uncertainties on $Y_{\mathrm{P}}, \, \Delta Y/\Delta Z$, and $(Z/X)_\odot$, the initial $Y$ and $Z$ abundances are obviously also affected by the observational error on [Fe/H]. Generally the errors quoted in the literature vary from about $\pm0.01$, probably underestimated, to $\pm0.1$. Here we adopt as a conservative error $\Delta \mathrm{[Fe/H]}=\pm0.05$.

By means of eqs. (\ref{eq:elio}) and (\ref{eq:metal}), taking the quoted uncertainties into account, we obtain eight values of ($Y$, $Z$) for each of the selected clusters, for which we compute pre-MS models. More precisely, the different $Y$ and $Z$ values are calculated by adopting, in turn, the minimum and the maximum values of one of the four parameters ([Fe/H], $Y_{\mathrm{P}}, \, \Delta Y/\Delta Z$, and $(Z/X)_\odot$), while the others are kept fixed to their central value. We computed two additional models with the maximum and minimum values of the estimated initial $^7$Li abundance, which, as already mentioned in Sect. \ref{sec:models}, is set to $\epsilon_{\mathrm{Li}} = 3.2\pm 0.2$. Obviously a change in the initial chemical composition also affects the position of the star in the colour-magnitude diagram, hence the age and mixing length parameter determination. This effect has been taken into account (see Sect. \ref{sec:ammassi}).

Another source of uncertainty related to the stellar chemical composition that has to be considered is the assumed distribution of heavy elements, at fixed Z, which strongly affects opacity values \citep{sestito06}. Since the opacity determines the temperature gradients and thus the extension of the convective envelope (in mass and temperature), a variation in this quantity, due to the uncertainty on the adopted mixture, modifies the lithium-burning rate and its resulting surface abundance.

Figure \ref{fig:misture} shows the relative differences among the \texttt{OPAL} radiative opacity tables\footnote{We use the \texttt{OPAL} opacity table released in 2005 for $\log T[K] >~4.5$, which are available at the url, \url{http://opalopacity.llnl.gov/opal.html}. For lower temperatures we use the \citet{ferguson05} radiative opacity, available at the url, \url{http://webs.wichita.edu/physics/opacity/}} computed by adopting the GN93, GS98, AS05 (our reference model), and AS09 solar mixtures. To make the figure much clearer , we also show a box representing the region covered by the entire convective envelope for stellar models with masses in the range 0.6 - 1.2 M$_\odot$ and [Fe/H] = $+0.0$.  The model ages have been chosen in the range 1 - 10 Myr, which roughly corresponds to the phase of efficient lithium burning for such masses.  

\begin{figure}[tp]
\centering
\includegraphics[width=\linewidth]{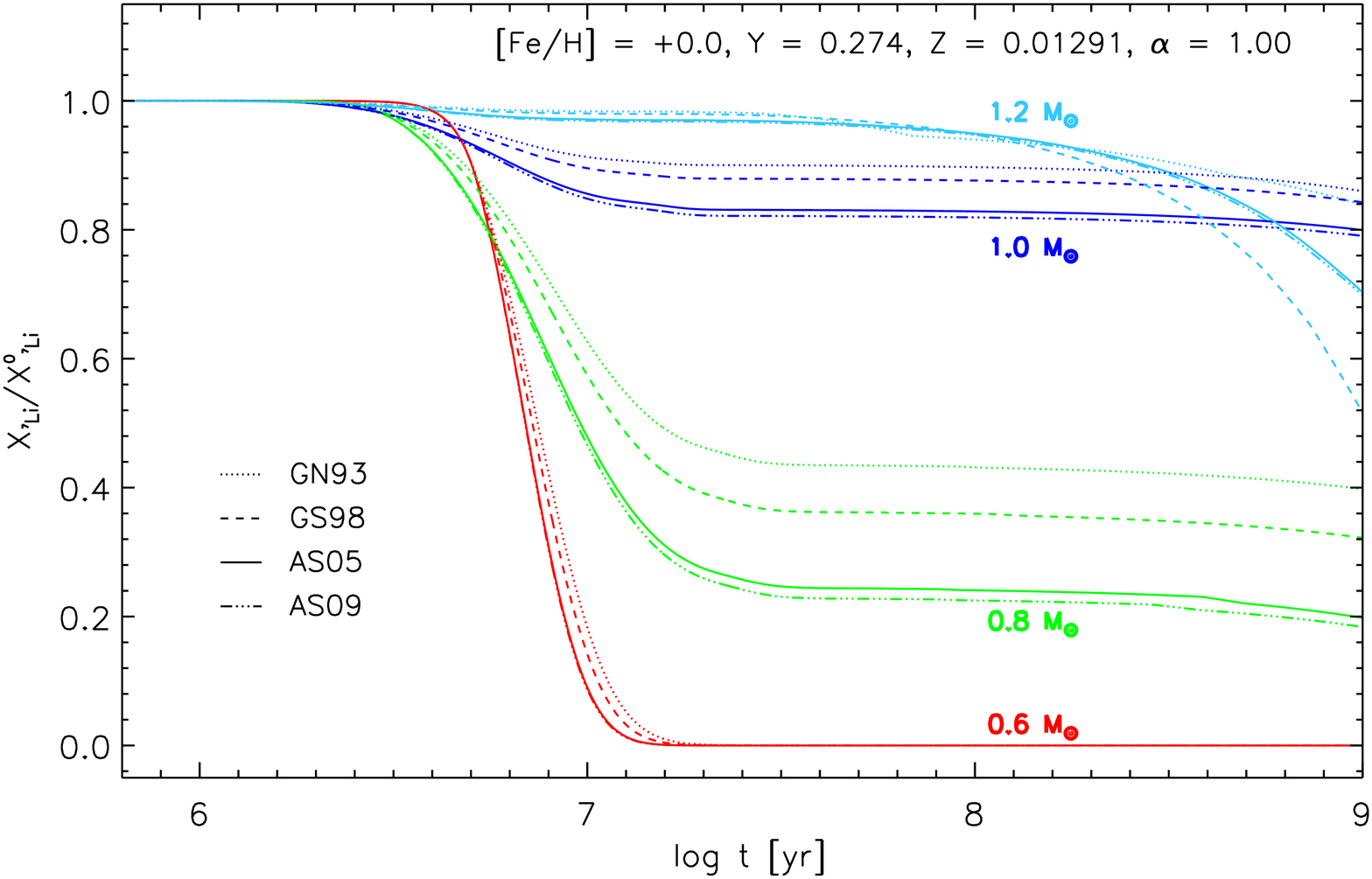}
\includegraphics[width=\columnwidth]{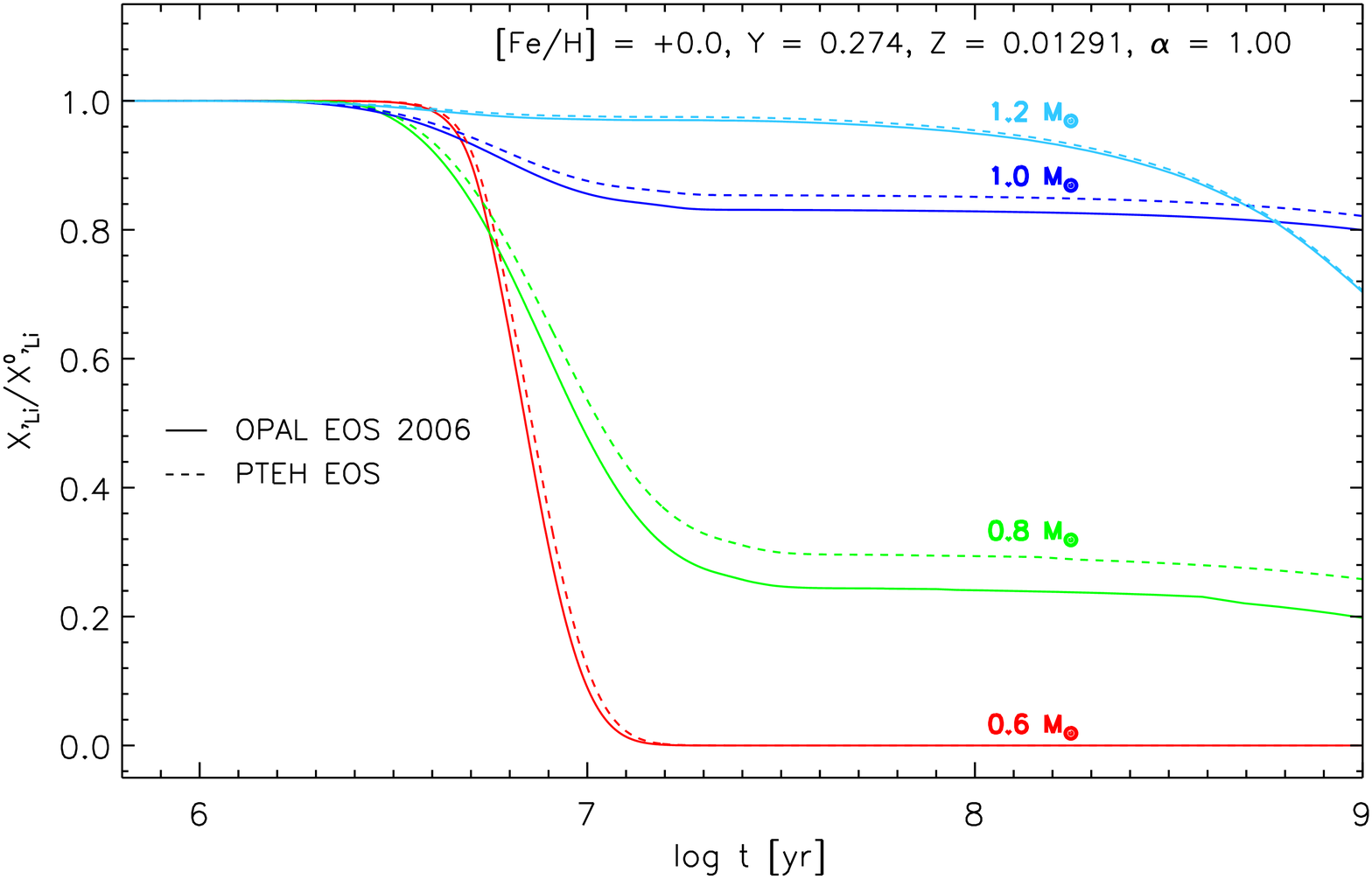}
\caption{Comparison among the surface lithium abundance obtained with our reference set of tracks (\textit{solid line}) and models with different assumptions on the adopted physical inputs, for M = 0.6, 0.8, 1.0, and 1.2 M$_\odot$ with $Z=0.01291$, $Y=0.274$, and $\alpha =1$. \textit{Upper panel}: effect of the change of the solar mixture (GN93, GS98, AS05, and AS09 ones) in the opacity tables. \textit{Bottom panel}: effect of adopting the \texttt{OPAL} 2006 and \texttt{PTEH} EOS.}
\label{fig:inputs1}
\end{figure}

The upper panel of Fig. \ref{fig:inputs1} shows the change in surface lithium abundance resulting from the adoption of the aforementioned solar mixtures in the opacity tables, in stars with different masses, namely 0.6, 0.8, 1.0, and 1.2 M$_\odot$. The increase of the iron group elements at fixed metallicity $Z$ leads to higher radiative opacity and in turn to a deeper convective envelope and consequently to greater lithium depletion. This is exactly what occurs when updating the heavy elements mixture to the recent AS05 or AS09 from the older GN93 or GS98. The higher the mass, the lower is the surface lithium depletion and, consequently, the sensitivity of the lithium burning to the opacity change. Since AS05 and AS09 metals distributions are quite similar, we expect negligible differences in the opacity coefficients, and thus in the model predictions. 

\subsection{Opacity coefficients.}

Besides the influence on the opacity of the heavy element distribution, it is worth analysing the error on the calculation of Rosseland radiative opacity coefficients $\kappa_\mathrm{R}$ at fixed chemical composition. The current version of the opacity tables we adopt in present calculations \citep[i.e. \texttt{OPAL} 2005 see e.g.,][]{rogers92,iglesias96} does not contain information about the related uncertainty. Thus to give a conservative uncertainty estimation on $\kappa_\mathrm{R}$, we evaluated the relative differences between the \texttt{OPAL} and the \texttt{OP} \citep[\emph{Opacity Project} see e.g,][]{seaton94,badnell05} radiative opacity coefficients in their full range of validity, once the same chemical composition has been adopted. We found that the maximum/minimum relative difference between the two opacity tables is close to $\pm 5\%$ in the region of interest for the present calculations (i.e. the convective envelope) \citep[see also,][]{neuforge01,badnell05,valle12}. Thus, we assume the value of $\Delta\kappa_\mathrm{R}/\kappa_\mathrm{R} = \pm5 \%$ as a conservative uncertainty.

\subsection{Equation of state}

Owing to the complexity of the evaluations of the various thermodynamical quantities, which are strictly correlated among each other, it is very difficult to assess a precise uncertainty on the EOS tables. An idea of how the current indetermination on the EOS propagates into stellar evolutionary predictions can be obtained by computing models with two different EOS tables that have been widely adopted, namely the \texttt{OPAL} EOS 2006 (our reference one) and \texttt{PTEH} \citep{pols95}\footnote{The \texttt{PTEH} EOS was computed using the FreeEOS fortran library developed by A.W. Irwin, which allows computing the EOS by the free-energy minimization technique. One of its advantages is the possibility of setting several flags to mimic other historical EOS.}. The comparison between the \texttt{OPAL} and the \texttt{PTEH} is useful to assess the effect of adopting a completely different treatment of the gas in stellar conditions, the two EOS being computed, respectively, in the formalism of the physical and chemical picture \citep[see e.g.,][]{trampedach06}.

The influence of the adopted EOS on the location in the HR diagram has already been discussed in several papers for both pre-MS evolution \citep[see e.g.,][]{mazzitelli89,dantona93,tognelli11} and low-mass MS stars \citep[see e.g.,][]{dorman89,neece84,chabrier97,dicriscienzo10}. Here, we simply recall that the models are particularly affected by the EOS in all the phases where a thick convective envelope is present, i.e. the pre-MS or MS structures of low- and very low-mass stars. In these phases, when lithium burning is efficient, the resulting surface $^7$Li abundance is quite sensitive to the adopted EOS, too, as shown in the bottom panel of  Fig. \ref{fig:inputs1}. As noticed above, surface lithium abundance gets less affected by the EOS change as the mass increases, because of the progressively reduction of the burning efficiency.

\subsection{$^7$Li(p,$\alpha$)$\alpha$ cross section}

Lithium destruction is obviously dependent on the $^7$Li(p,$\alpha$)$\alpha$ cross section. However, the current uncertainty on the quoted reaction rate for bare nuclei is quite small \citep[a few percent see e.g.,][]{nacre,lattuada01}, so that the effect on $^7$Li abundance of such error is very small compared to the other error sources. We adopt the value of $\pm 5\%$ as a conservative uncertainty on this quantity.

\subsection{Total uncertainty on $^7$Li surface abundance predictions}
\label{sec:total}

The partial uncertainty due to each parameter/physical input was obtained by the difference between the reference model, which is the one computed with the reference values of all the parameters, and the model computed by varying such parameter. This procedure was iterated for all the uncertainty sources discussed in the text. Then, the total error on surface lithium abundance predictions was computed by quadratically adding all the partial errors. We want to emphasize that the uncertainty analysis was performed for all the chemical compositions suitable for the selected clusters. Thus, for each cluster, error bars consistent with its chemical composition, mixing length parameter, and age were evaluated.

\section{Surface lithium abundance: theory vs observations}
\label{sec:results}

\subsection{Young open clusters}
\label{sec:ammassi}
%
\begin{table}[t]
\caption{Main properties adopted/derived for the five selected open clusters.}
\label{tab:oc_cmd}
\centering
\scriptsize
\begin{tabular}{l|cccccc}
\hline
Cluster: & [Fe/H]: & ($Y$, $Z$): & $\alpha_{\mathrm{MS}}$ & age (Myr): & age (Myr): \\
	         &            &                    &                                     & ($\lambda_\mathrm{ov} = 0.0$) &  ($\lambda_\mathrm{ov} = 0.2$)\\
\hline
\hline
\\
Ic2602 & $+0.00$\,\tablefootmark{a} & ($0.274$, $0.0129$) & $1.68 \pm 0.1$ & $40\pm10$ & $55\pm10$\\
\\
$\alpha$ Per &  $-0.10$\,\tablefootmark{b} & ($0.269$, $0.0104$) & $1.68 \pm 0.1$ & $60\pm10$ & $75\pm10$\\
\\
Blanco1 & $+0.04$\,\tablefootmark{c} & ($0.276$, $0.0141$) & $1.90^{+0.1}_{-0.2}$ & $110\pm30$ & $130\pm30$\\
\\
Pleiades & $+0.03$\,\tablefootmark{d} & ($0.276$, $0.0138$) & $1.90^{+0.1}_{-0.2}$ & $120\pm20$ & $130\pm20$\\
\\
Ngc2516 & $-0.10$\,\tablefootmark{e} & ($0.269$, $0.0104$) & $1.90 \pm 0.1$ & $130\pm20$ & $145\pm20$\\
               & $+0.07$\,\tablefootmark{f} & ($0.278$, $0.0150$) & $1.90 \pm 0.1$ & $130\pm20$ & $145\pm20$\\
\\
\hline
\end{tabular}
\tablefoot{The columns list, respectively, the cluster's name, [Fe/H], initial helium and metal abundance ($Y$, $Z$), the mixing length parameter calibrated on MS stars ($\alpha_\mathrm{MS}$), the best fit age without core overshooting, and the best fit age with a core overshooting parameter set to $\lambda_\mathrm{ov} = 0.2$.}
\tablebib{\tablefoottext{a}{\citet{dorazi09}};
	\tablefoottext{b}{\citet{balachandran11}};
	\tablefoottext{c}{\citet{ford05}};
	\tablefoottext{d}{\citet{soderblom09}}; 
	\tablefoottext{e}{\citet{sung02}};
	\tablefoottext{f}{\citet{terndrup02}}.}
\end{table}

The clusters age and the mixing length parameters for MS stars are determined by comparing the observed CMDs with the present theoretical isochrones. The age is largely affected by the lack of stars near the overall contraction region of such young clusters, and it is only marginally affected by the uncertainty on the chemical composition adopted for the calculations.  Similarly, the uncertainty on the calibration of $\alpha_\mathrm{MS}$ essentially comes from the spread of the MS in the CMD. The effects of the indetermination on age and $\alpha_\mathrm{MS}$ on surface $^7$Li abundance are evaluated for each cluster and quadratically added to the other error sources to define the theoretical surface lithium abundance error bars. Table \ref{tab:oc_cmd} summarizes the results for the main parameters of each cluster, namely, chemical composition, $\alpha_{\mathrm{MS}}$, and age. For Ngc 2516 two different [Fe/H] values are available: (1)~[Fe/H] = $-0.10$ (photometric) and (2) [Fe/H] = $+0.07$ (spectroscopic). Since the spectroscopic value is still quite uncertain, because it has been determined using only two stars \citep[see the discussion in ][]{terndrup02}, we decided to also use the photometric one for the following comparison with surface lithium data. 

The age determination for the selected clusters was performed both with models without core overshooting ($\lambda_\mathrm{ov} = 0$) and with a standard upper limit for the presence of overshooting, $\lambda_\mathrm{ov} = 0.2$ \citep[see e.g.,][and references therein]{brocato03,claret07}, in order to take a suitable range of $\lambda_\mathrm{ov}$ into account. However, as shown in Table \ref{tab:oc_cmd}, the age difference is for all the cases close to the quoted uncertainty. Moreover, the impact of the age uncertainty on $\epsilon_\mathrm{Li}-T_\mathrm{eff}$ profile is almost negligible for clusters older than about 80 - 90 Myr, since all the stars have already reached the ZAMS, so that their position in the CMD is weakly dependent on the age. The situation is different for very young clusters, such as Ic2602 and $\alpha$ Per. In this case low-mass stars are close but not yet in ZAMS, so an age variation produces an appreciable change in $T_\mathrm{eff}$.

As for the age, the $\alpha_\mathrm{MS}$ uncertainty also has a small effect on the predictions of surface $^7$Li, since in the age and mass range we are investigating, the stars undergo an almost negligible depletion during MS. Only the surface $^7$Li abundances for the lowest masses of the sample, i.e. M $\la$ 0.6 M$_\odot$, are marginally affected by such a variation in the predicted $^7$Li. Besides this, $\alpha_\mathrm{MS}$ slightly changes the effective temperature of the stars, which, however, is a second-order effect when compared to the $T_\mathrm{eff}$ shift introduced by the other uncertainties previously discussed.

\begin{figure*}[t]
\centering
\hspace{-0.5cm}
\includegraphics[width=\columnwidth]{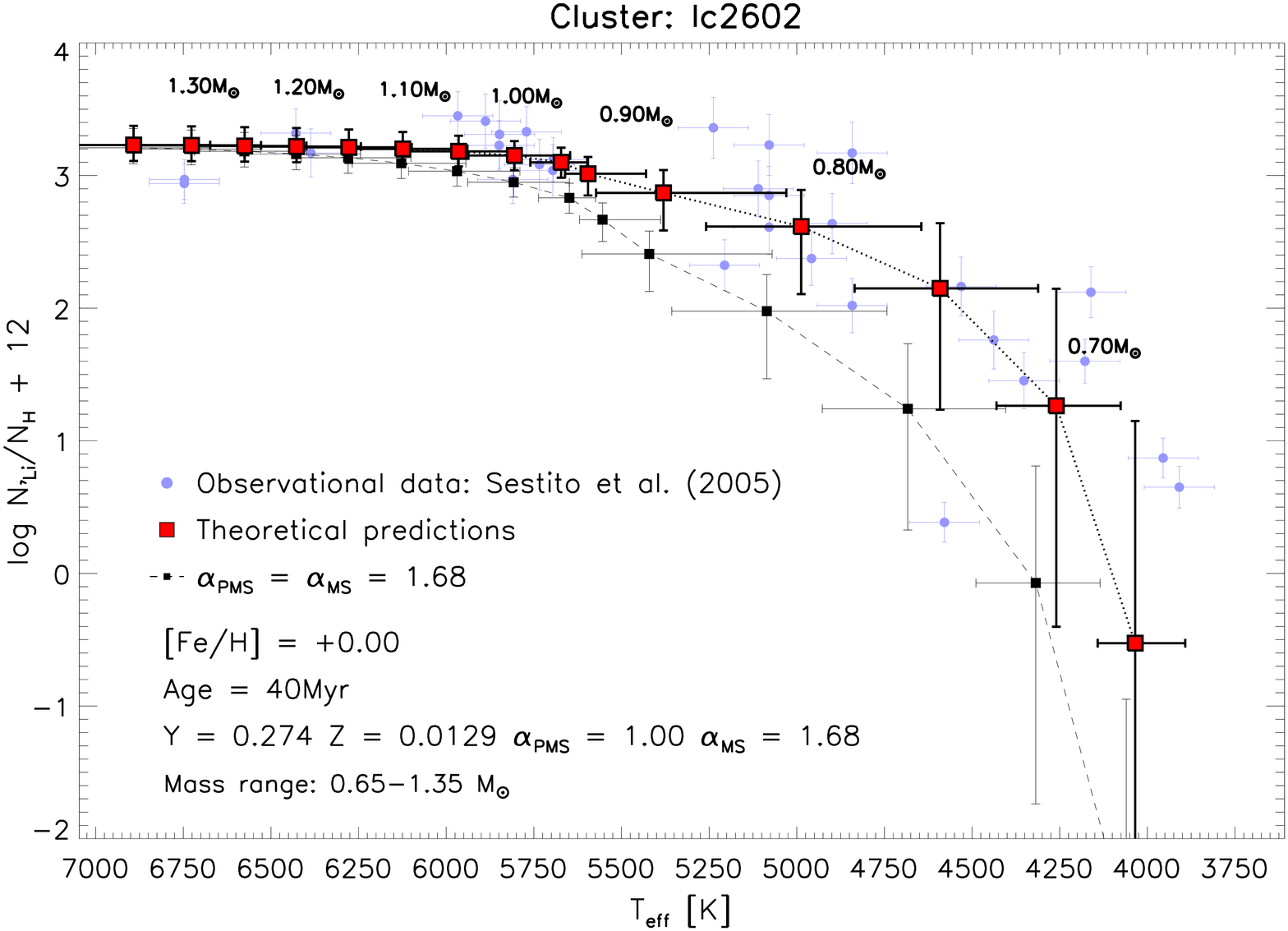}
\includegraphics[width=\columnwidth]{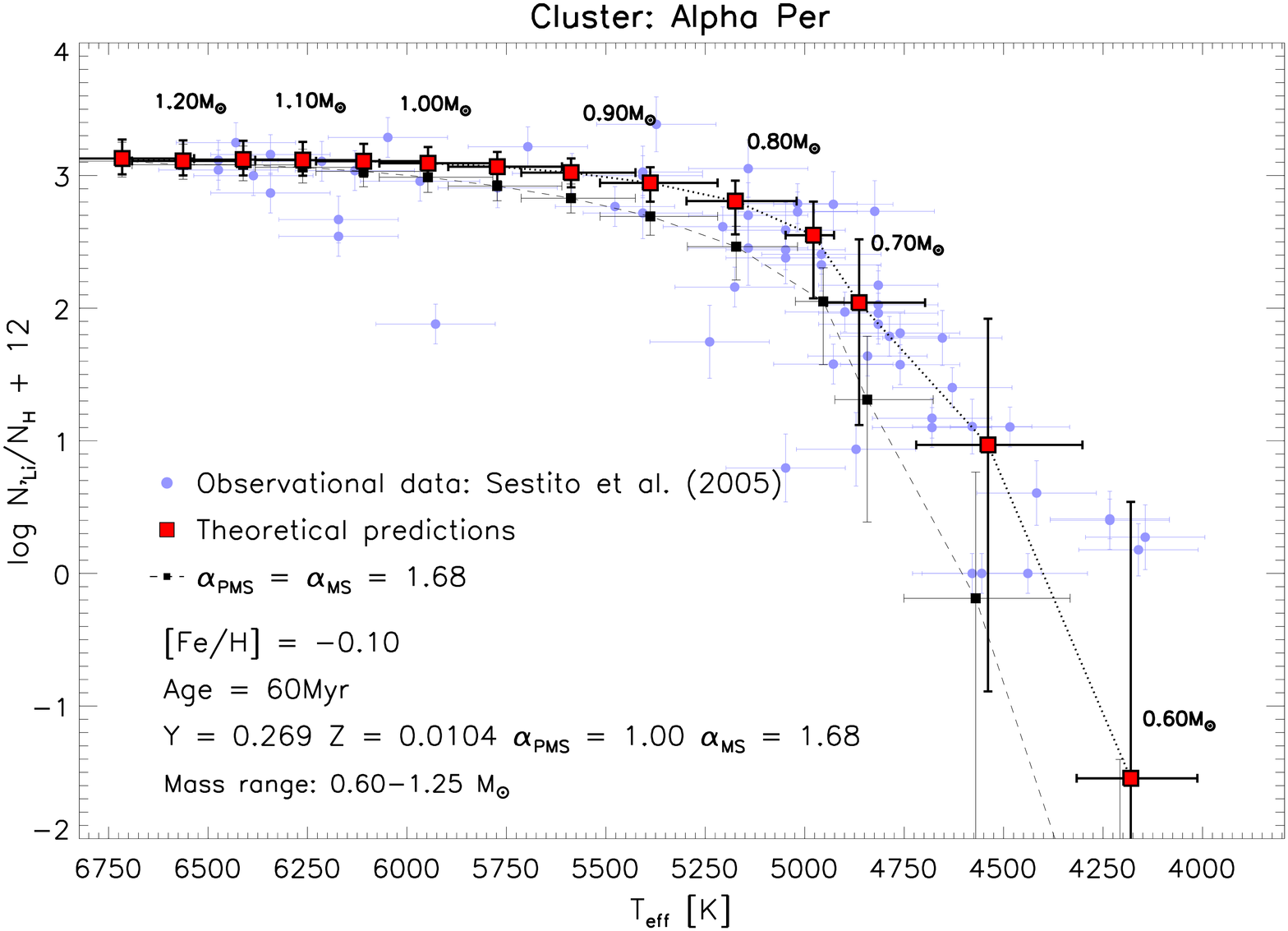}\\
\vspace{0.1cm}
\hspace{-0.5cm}
\includegraphics[width=\columnwidth]{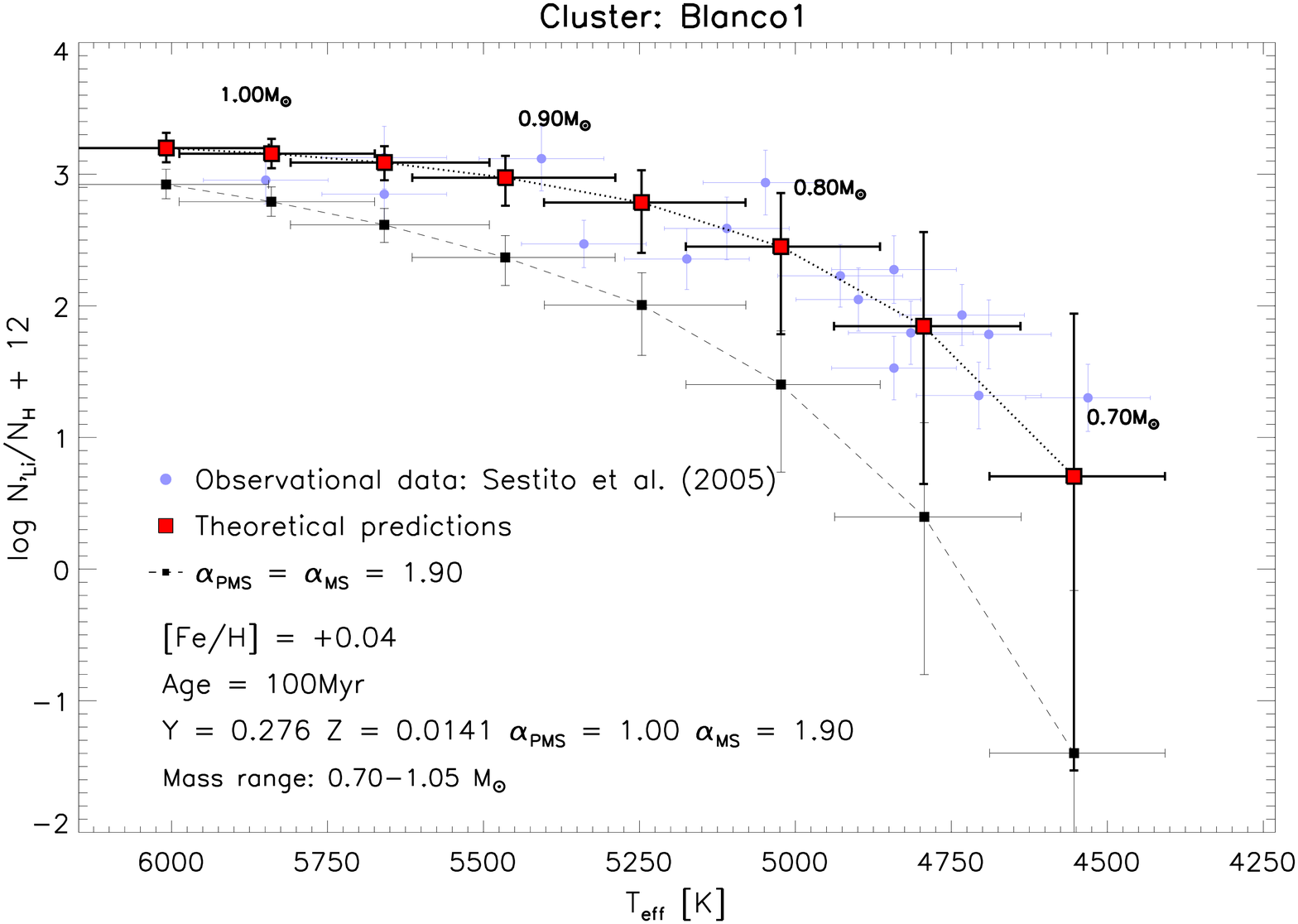}
\includegraphics[width=\columnwidth]{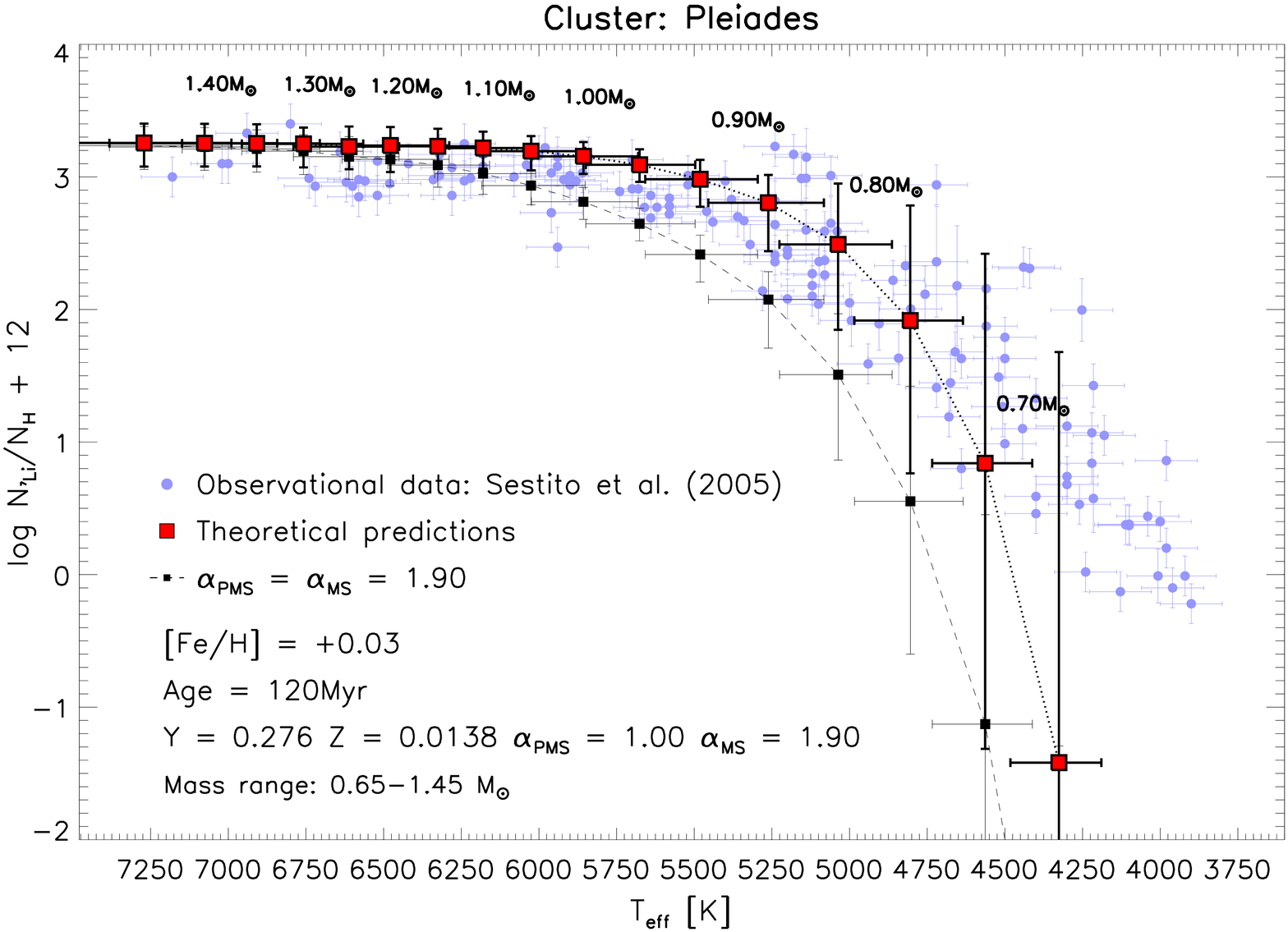}\\
\vspace{0.1cm}
\hspace{-0.5cm}
\includegraphics[width=\columnwidth]{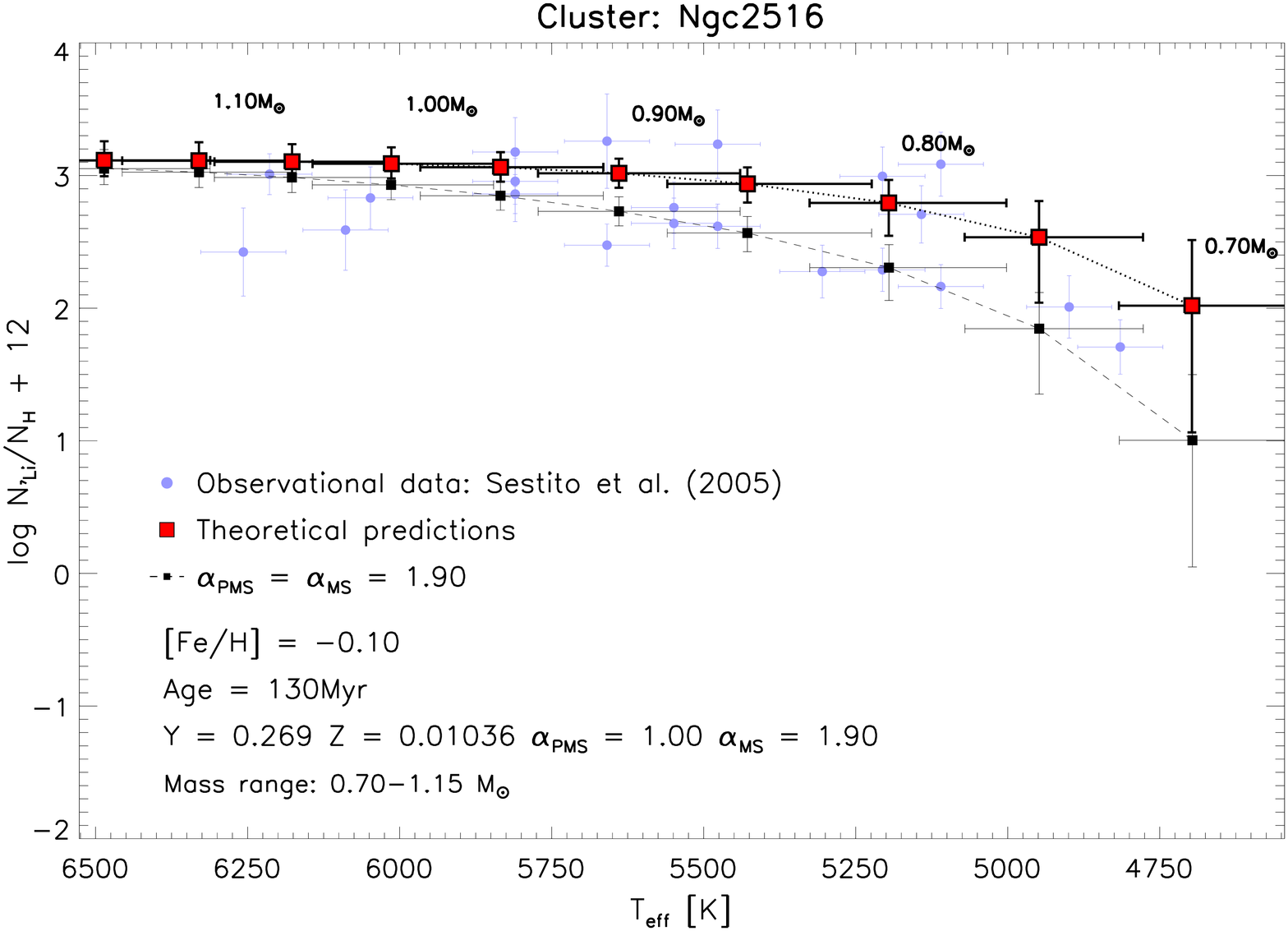}
\includegraphics[width=\columnwidth]{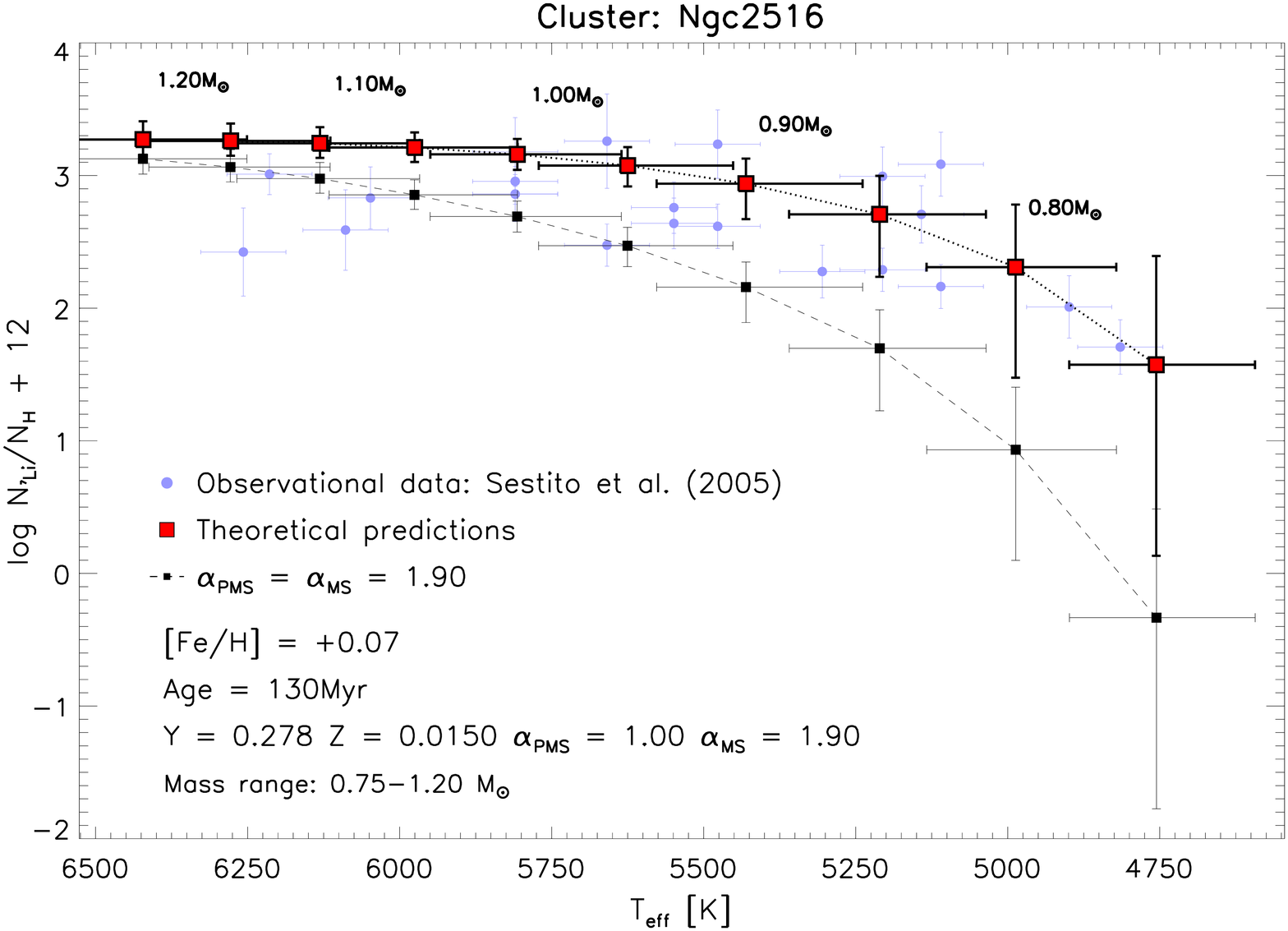}
\caption{Comparison between our model predictions and observational data (\textit{filled circles}) for surface lithium abundance, for the selected sample of young open clusters, namely (from the top left panel), Ic2602, $\alpha$ Per, Blanco 1, Pleiades, and Ngc 2516. In each panel we show both the low-convection efficiency  ($\alpha_\mathrm{PMS}=1.0$, \textit{red-filled squares} and \textit{dotted line}) and the high-convection efficiency models ($\alpha_\mathrm{PMS} = \alpha_\mathrm{MS}$, \emph{dashed line} and \emph{small black-filled squares}). We also plotted the mass and the corresponding theoretical uncertainties on both the effective temperature and lithium abundance, for low- and high-convection efficiency models.}
\label{fig:litio}
\end{figure*}

Regarding the comparison between surface $^7$Li data and present models, as a first step, we make the assumption of a constant value of the mixing length parameter from the early pre-MS phase to the MS. Figure \ref{fig:litio} shows the comparison between the theoretical predictions and data for each cluster. As one can see, the models with $\alpha_\mathrm{PMS}=\alpha_\mathrm{MS}$ (dashed line with filled black squares) fail in reproducing the observed $^7$Li abundances in almost all the selected clusters, for stars less massive than about 1 M$_\odot$, even if theoretical and observational errors are taken into account.  For these stars, the predicted $^7$Li is systematically lower than what is observed, with differences as great as 1 dex for low-mass stars (about 0.6 - 0.7 M$_\odot$), confirming  the well known disagreement between theory and observations for $^7$Li surface abundance, even in light of the recent data and taking the updated theoretical models into account within present error estimates.

Models computed with $\alpha_\mathrm{PMS}=\alpha_\mathrm{MS}$ partially agree with data only in the case of $\alpha$ Per for M $\ga 0.7$ M$_\odot$, and Ngc 2516 if the low photometric [Fe/H] value is adopted (bottom left panel of Fig. \ref{fig:litio}). If the spectroscopic [Fe/H] value is used for NGC 2516, the predictions, as the other clusters, do not match the observations for M $\la 1$ M$_\odot$. However, we emphasize that, for these two clusters, models and data are compatible each other because of the large $^7$Li abundance scatter present among stars with similar $T_\mathrm{eff}$ (about 1 dex), combined with the large error bars on theoretical predictions.

Since in most of the cases the models with $\alpha_\mathrm{PMS}=\alpha_\mathrm{MS}$ disagree with the data, and given the high sensitivity of $^7$Li surface abundance predictions to the convection efficiency, it is worth exploring the possibility that the mixing length parameter value varies from the pre-MS to the MS phases. Indeed, a possible dependence of $\alpha$ on the evolutionary phase (and/or gravity, $T_\mathrm{eff}$, mass) is suggested from both observations and hydrodynamical simulations, as discussed in the introduction. Thus, we computed models with different values of $\alpha_\mathrm{PMS}$, namely, $\alpha_\mathrm{PMS} = 1.0$, 1.2, 1.4, 1.68, and 1.9, once $\alpha_\mathrm{MS}$ and the ages have been fixed by the comparison in the CMD. Figure \ref{fig:litio} shows the comparison between our `best fit' models and $^7$Li data for each cluster (dotted lines and filled red squares). The theoretical error bars computed for each cluster are also shown. 

We emphasize that a satisfactory agreement with all the clusters in the sample (with the exception of the Pleiades) can be achieved by assuming the same pre-MS convection efficiency, namely $\alpha_\mathrm{PMS}=1.0$. Such low-convection efficiency models are able to reproduce, within the error bars, the mean depletion profile even for low-mass stars, especially in the case of Ic 2602 and $\alpha$ Per. 

As shown in Fig. \ref{fig:litio}, the poorest match between theory and data is achieved for the Pleiades. The hottest stars are nearly compatible within the error bars with the observations, which show a surface abundance about 0.2 - 0.3 dex lower than the predicted one. A possible way to improve the agreement with these stars is to adopt an initial lithium abundance of about $\epsilon_\mathrm{Li} \approx 3$. However, this method does not improve the agreement with the low-mass stars, a problem still largely discussed in the literature \citep[see e.g.][and references therein]{king00,jeffries00,umezu00,dantona03,clarke04,xiong06,king10}. 

The results we obtain for young open clusters confirm the partial results of previous analysis, which have noticed that models with low-convection efficiency during pre-MS phase agree much better with lithium observation than those with solar or MS calibrated values \citep[see e.g.,][]{ventura98,dantona03,landin06}. 

\begin{figure*}[t]
\centering
\includegraphics[width=0.98\columnwidth]{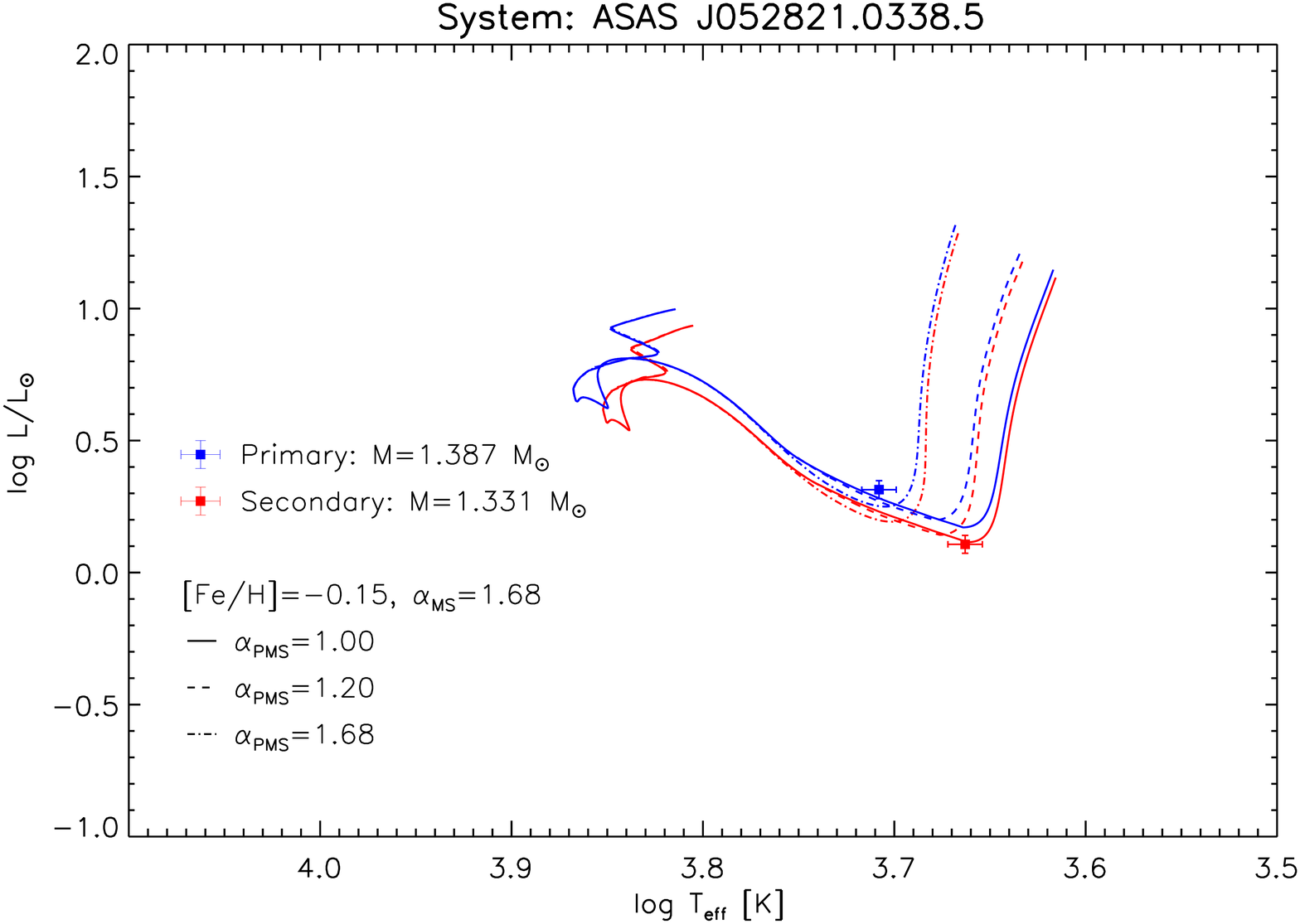}
\includegraphics[width=0.98\columnwidth]{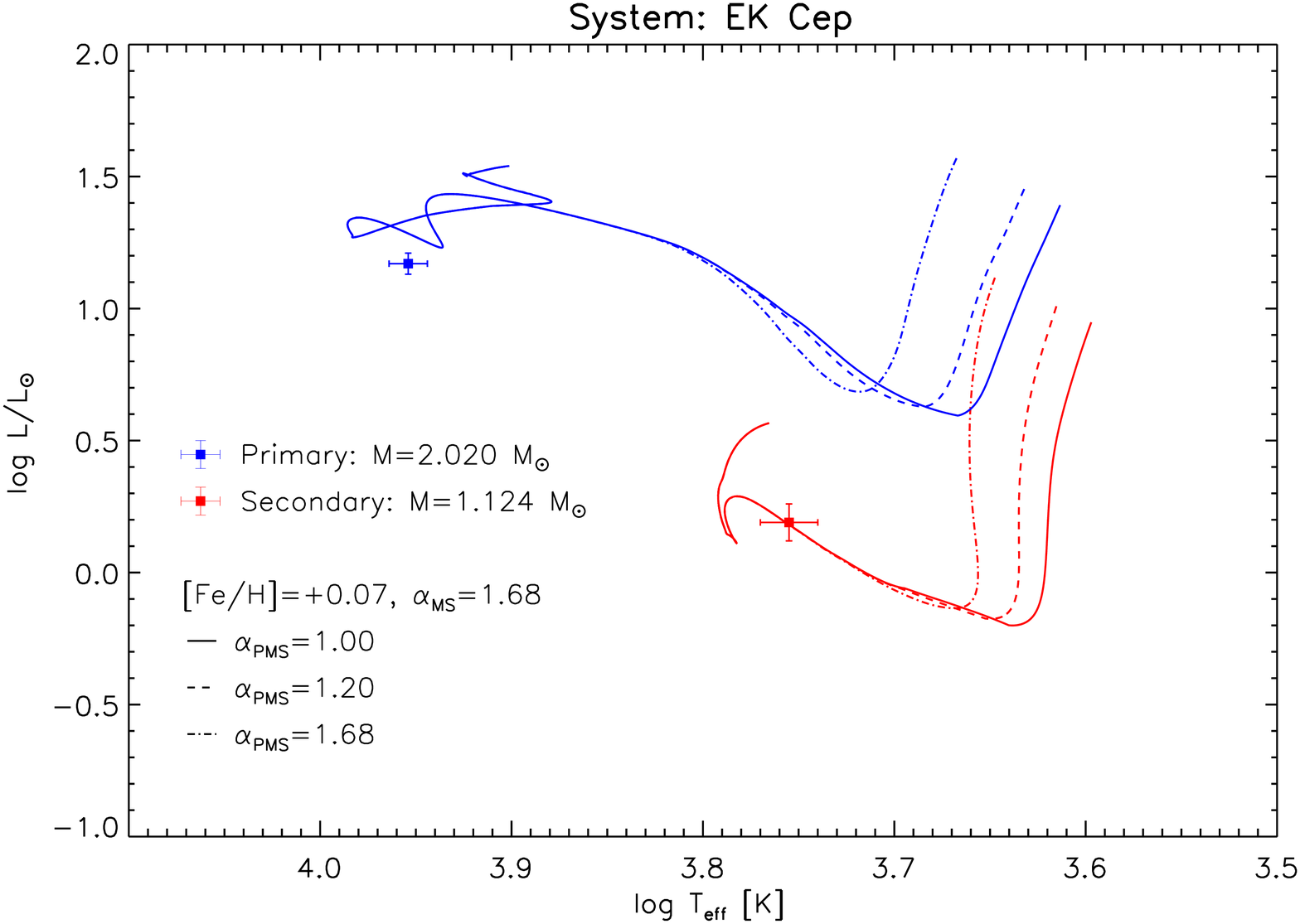}
\includegraphics[width=0.98\columnwidth]{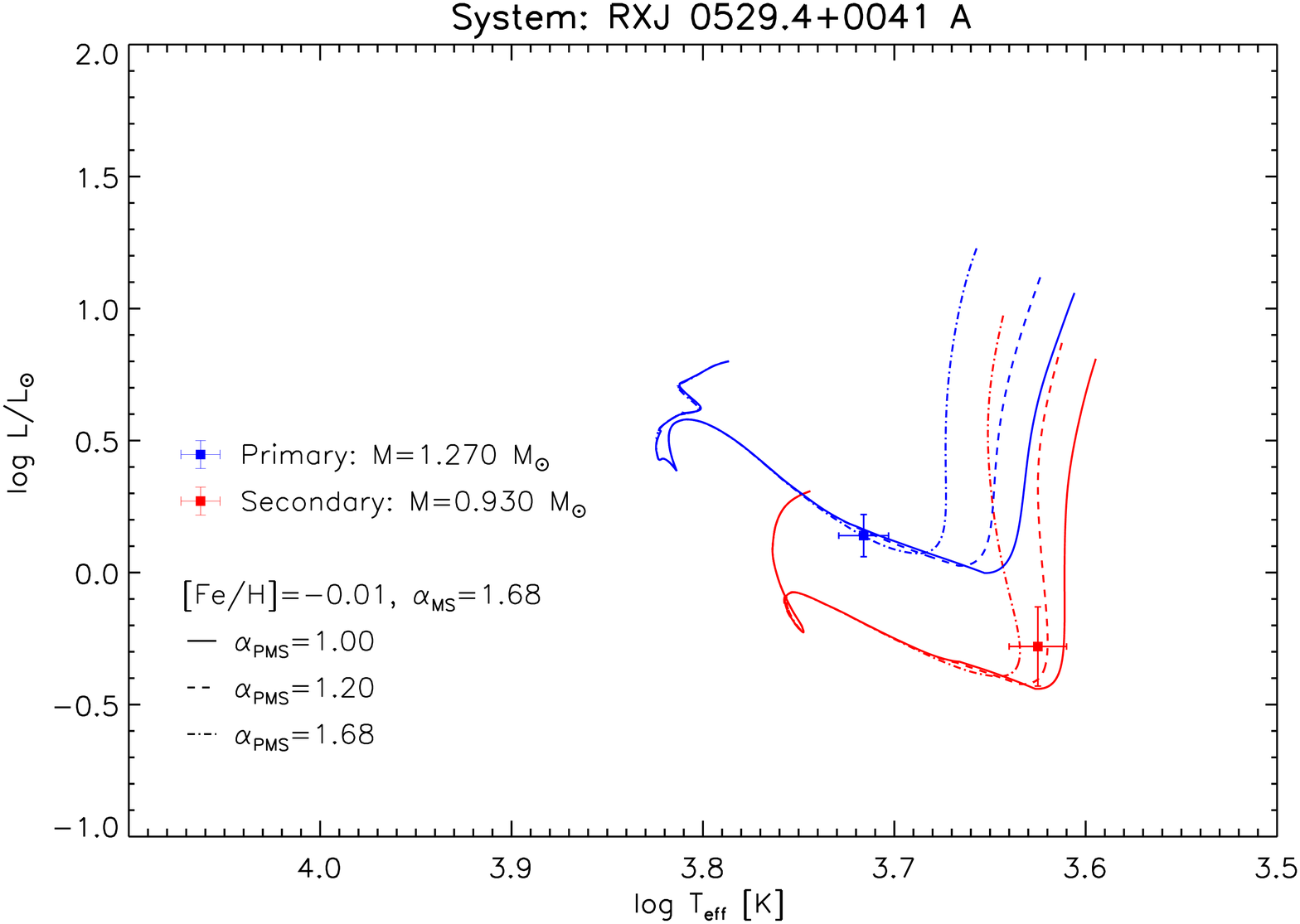}
\includegraphics[width=0.98\columnwidth]{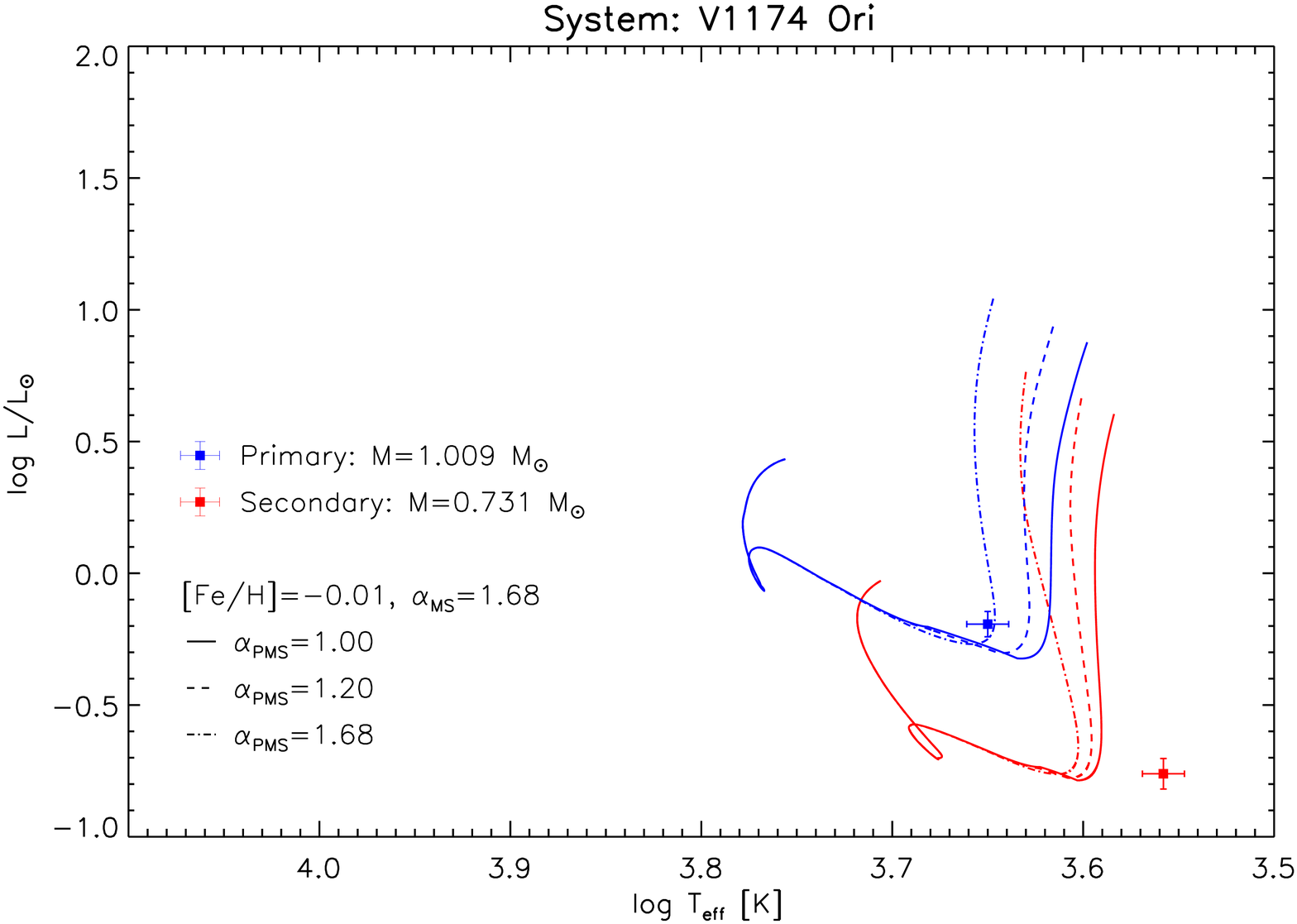}
\caption{HR diagram for the selected sample of EBs (see text). For each system we plotted the models for the labelled chemical composition and mass, corresponding to the primary (\textit{blue line}) and the secondary star (\textit{red line}). Models have been computed by adopting three different values for the mixing length parameter, namely $\alpha_\mathrm{PMS} = 1.0$ (\textit{solid line}), 1.20 (\textit{dashed line}), and 1.68 (\textit{dot-dashed line}). Observational data are shown (\textit{filled circles}), together with their uncertainty.}
\label{fig:hr_bin}
\end{figure*}
%
\begin{figure*}[t]
\centering
\includegraphics[width=0.98\columnwidth]{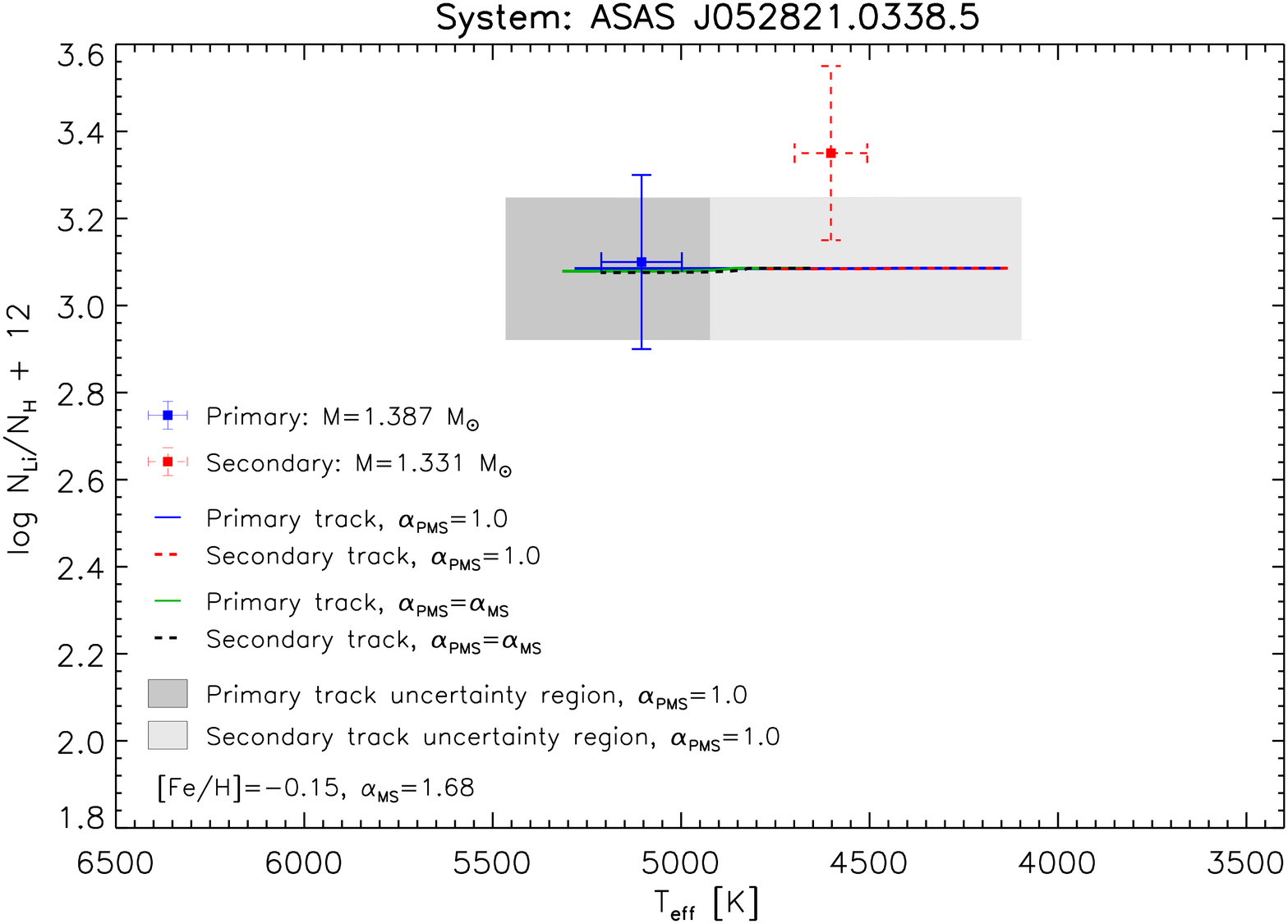}
\includegraphics[width=0.98\columnwidth]{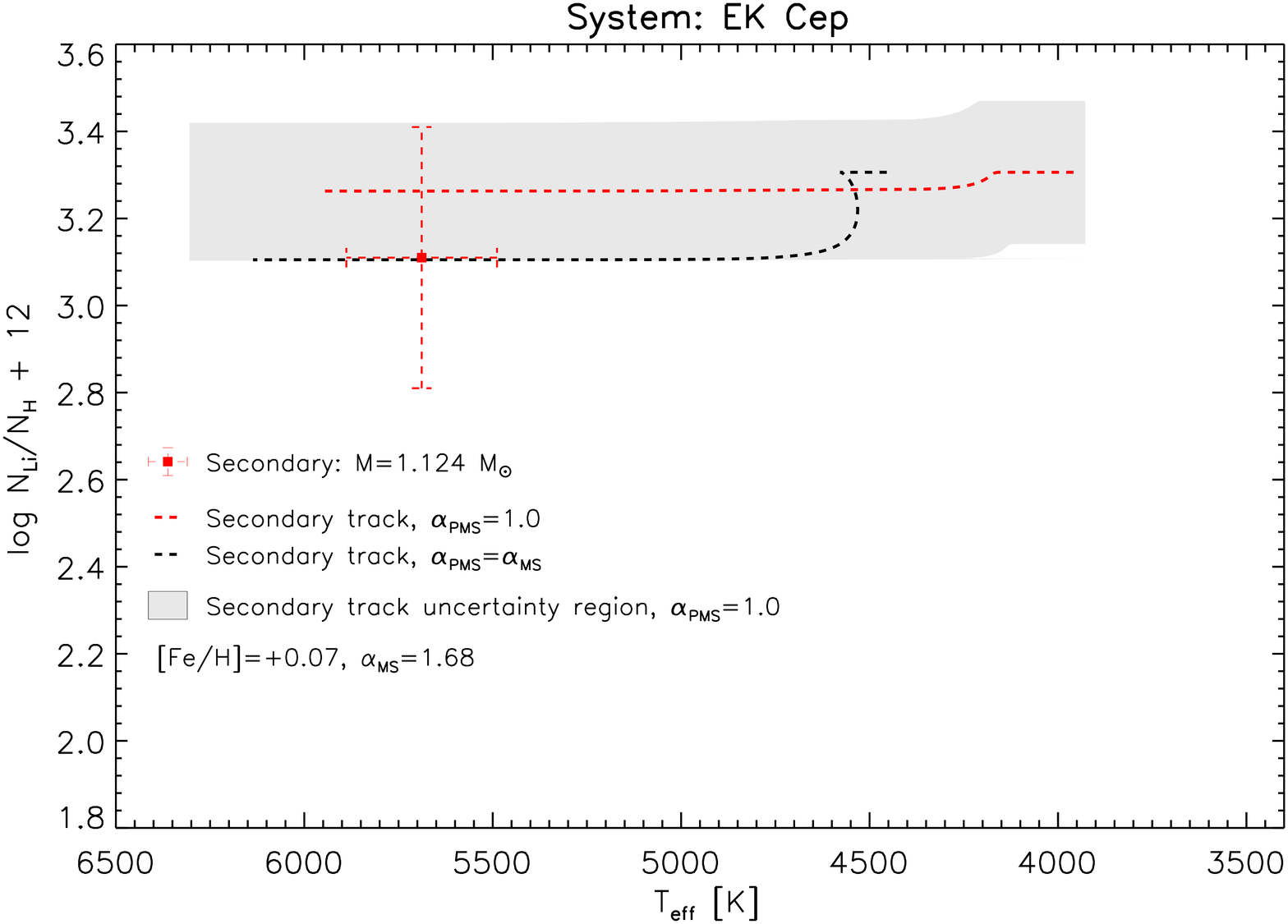}
\includegraphics[width=0.98\columnwidth]{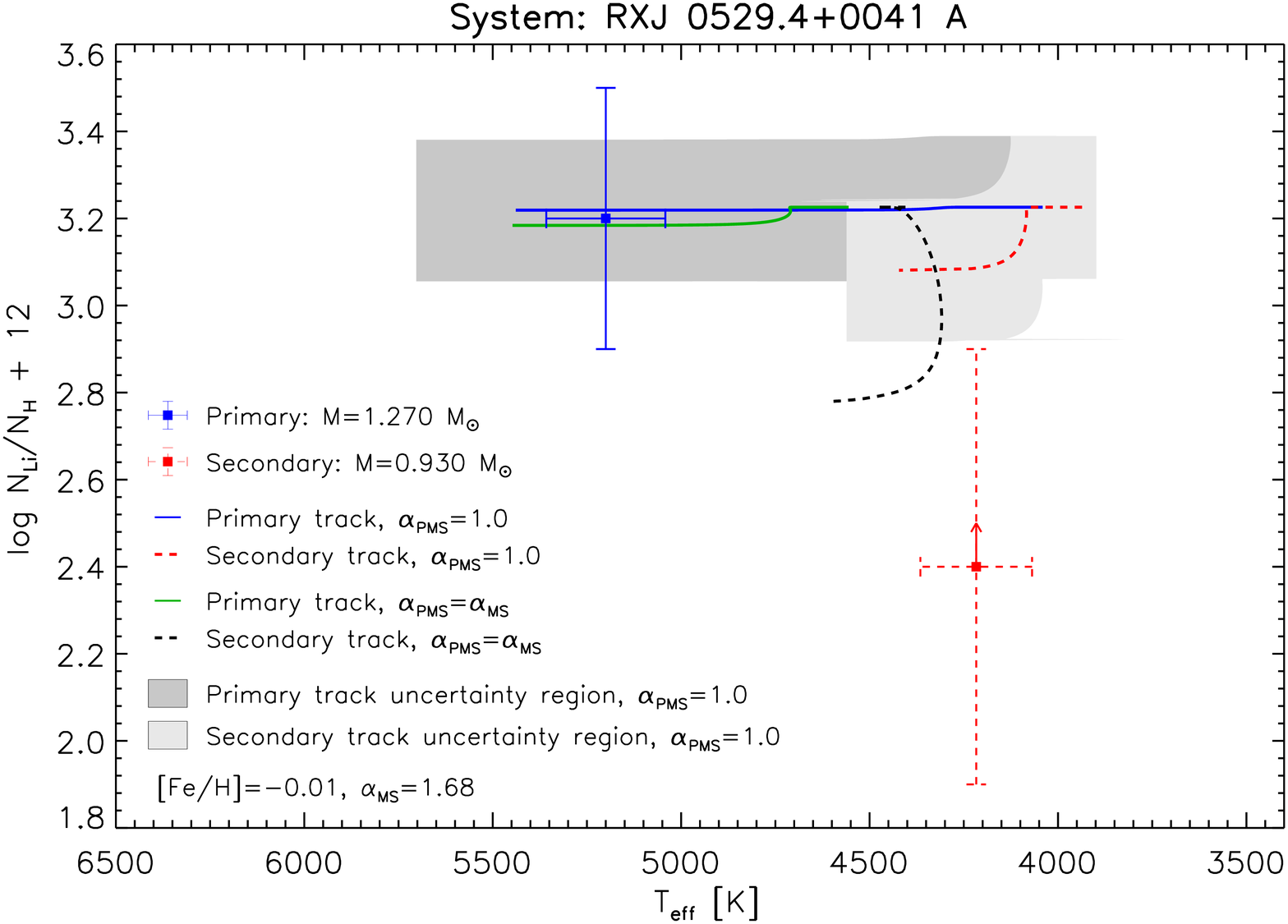}
\includegraphics[width=0.98\columnwidth]{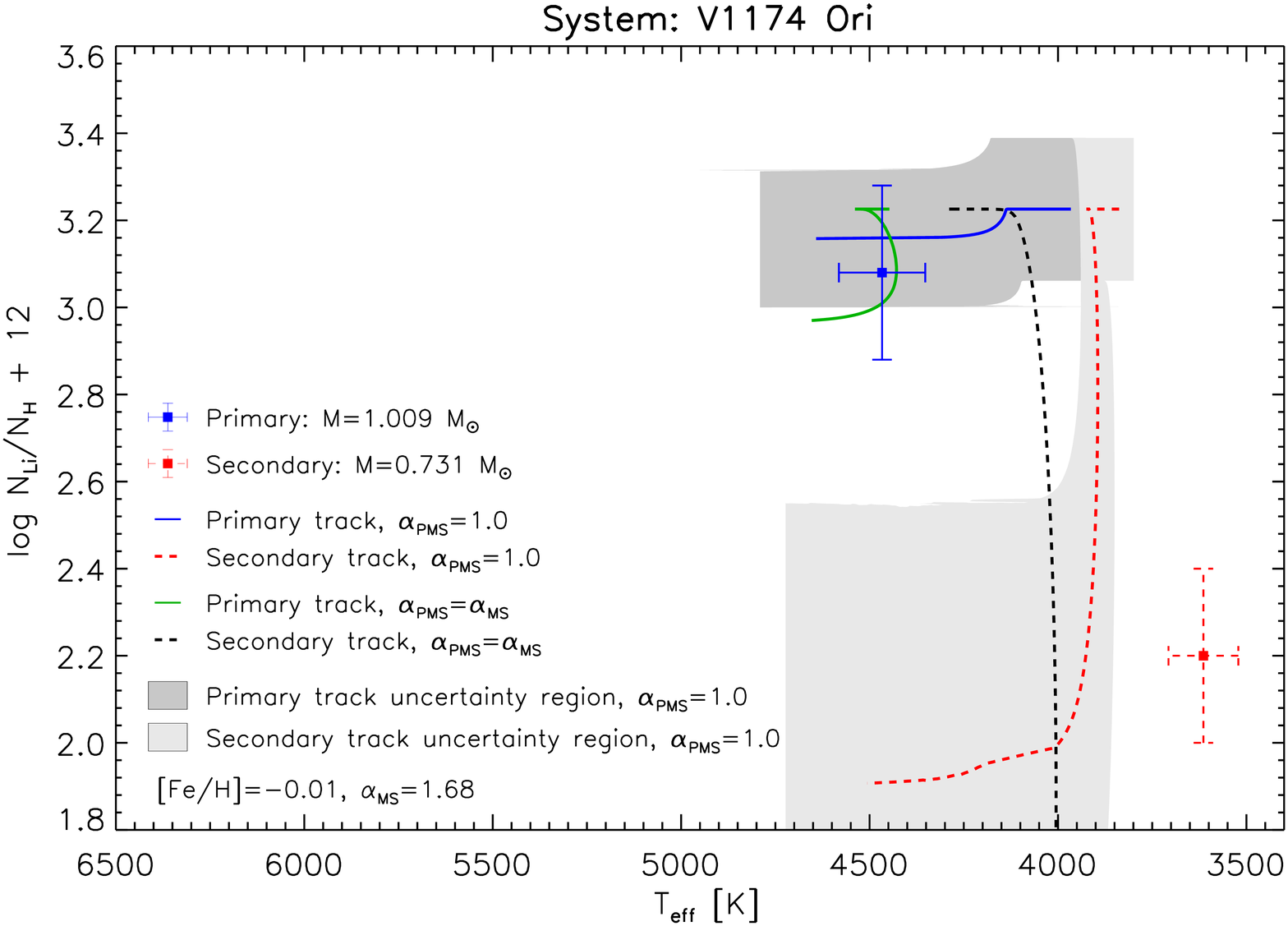}
\caption{Comparison between theoretical predictions and observations for surface lithium abundance in the selected  binary systems. Lithium evolutionary tracks for low-convection efficiency ($\alpha_\mathrm{PMS} =1.0$) corresponding to the primary (\textit{blue-solid line}) and the secondary star (\textit{red-dashed line}) have been computed for the labelled chemical composition and mass. Shaded area represents the uncertainty regions for the primary (\textit{dark-shaded area}) and secondary stars in the system (\textit{light-shaded area}). We also show the primary and secondary evolutionary tracks computed with $\alpha_\mathrm{PMS}=\alpha_\mathrm{MS}$ (\emph{green-solid} and \emph{black-dashed lines}, respectively).}
\label{fig:lit_bin}
\end{figure*}
\subsection{Binary stars}
\label{sec:binary}
%
\begin{table*}
\centering
\caption{Main parameters adopted for the selected set of EBs stars.}
\footnotesize
\label{tab:binarie}
\begin{tabular}{lccccc}
\hline
\hline
System & Mass [M$_\odot$] & $\log T_\mathrm{eff} \mathrm{[K]}$ & $\log L/L_{\odot}$ & $\epsilon_\mathrm{Li}$ & [Fe/H]\\
\hline
\\
ASAS J052821+0338.5 (a) & $1.387 \pm0.017$ & $3.708\pm0.009$ & $0.314\pm0.034$ & $3.10\pm0.20$\,\tablefootmark{a} & $-0.20\pm0.20$\,\tablefootmark{e}\\
ASAS J052821+0338.5 (b)  & $1.331 \pm0.017$ & $3.663\pm0.009$ & $0.107\pm0.034$ & $3.35\pm0.20$\,\tablefootmark{a} & $-0.10\pm0.20$\,\tablefootmark{e} \\
\\
EK Cep (a) & $2.020\pm0.010$ & $3.954\pm0.010$ & $1.170\pm0.040$ & $-$ & $+0.07\pm0.05$\,\tablefootmark{f} \\ 
EK Cep (b) & $1.124\pm0.012$ & $3.755\pm0.015$ & $0.190\pm0.070$ & $3.11\pm0.30$\,\tablefootmark{b} & $+0.07\pm0.05$\,\tablefootmark{f} \\
\\
RXJ 0529.4+0041 A (a) & $1.270\pm0.010$ & $3.716\pm0.013$ & $0.140\pm0.080$ & $3.20\pm0.30$\,\tablefootmark{c} & $-0.01\pm0.04$\,\tablefootmark{g} \\
RXJ 0529.4+0041 A (b) & $0.930\pm0.010$ & $3.625\pm0.015$ & $-0.280\pm0.150$ & $2.40\pm0.50$\,\tablefootmark{c} & $-0.01\pm0.04$\,\tablefootmark{g} \\
\\
V1174 Ori (a) & $1.009\pm0.015$ & $3.650\pm0.011$ & $-0.193\pm0.048$ & $3.08\pm0.20$\,\tablefootmark{d} & $-0.01\pm0.04$\,\tablefootmark{g} \\
V1174 Ori (b) & $0.731\pm0.008$ & $3.558\pm0.011$ & $-0.761\pm0.058$ & $2.20\pm0.20$\,\tablefootmark{d} & $-0.01\pm0.04$\,\tablefootmark{g} \\
\\
\hline
\end{tabular}
\normalsize
\tablefoot{For ASAS J052821.0338.5 (a) and (b) components we used the averaged value of [Fe/H] = $-0.15$ when computing the models.}
\tablebib{\tablefoottext{a}{\citet{stempels08}};
	\tablefoottext{b}{\citet{martin93}};
	\tablefoottext{c}{\citet{covino00}};
	\tablefoottext{d}{\citet{stassun04}}; 
	\tablefoottext{e}{\citet{stempels08}};
	\tablefoottext{f}{\citet{martin93}};
	\tablefoottext{g}{\citet{dorazi09b}}.}
\end{table*}
Binary systems and, in particular, the subclass of detached double-lined eclipsing binaries (EBs) are severe tests for stellar models. Indeed, for EBs independent measurements of mass, radius, and effective temperature are available \citep[for a detailed review see e.g.,][]{mathieu07}.

The validity of our theoretical models have already been tested against a large sample of pre-MS binaries (26 objects) by \citet{gennaro11}, using the models of the  Pisa pre-MS database against observations by means of a Bayesian method. The present pre-MS models differ from those available in the quoted database only in the minimum value of the mixing length parameter, i.e. $\alpha = 1.0$ instead of $\alpha = 1.2$.

From the sample of EBs presented in \citet{gennaro11}, we selected a subsample of binary systems for which surface lithium abundances are available, namely ASAS~J052821+0338.5 \citep{stempels08}, EK Cep \citep{popper87}, RXJ 0529.4+0041 A \citep{covino04}, and V1174 Ori \citep{stassun04}. Table \ref{tab:binarie} summarizes the main parameters of each system: mass, effective temperature, luminosity, lithium abundance, and [Fe/H]. The corresponding models are computed for $\alpha_\mathrm{PMS}=1.0$, 1.2, and 1.68.

Figure \ref{fig:hr_bin} shows the HR diagram of the four selected systems compared with our evolutionary tracks. \citet{gennaro11} have already shown that theoretical models with low initial helium abundance and mixing length parameter agree better with the data of pre-MS binary systems, in particular for those ones with at least one component near the Hayashi track. In our sample only EK Cep does not have stars near the Hayashi track.

For ASAS J052821+0338.5 (Fig. \ref{fig:hr_bin}), both the lowest and the highest $\alpha_\mathrm{PMS}$ values are compatible with the primary star, whereas $\alpha_\mathrm{PMS} = 1.0$ - 1.2 is required to match the secondary. For Ek Cep we cannot constrain the mixing length value during the pre-MS since both stars are approaching the ZAMS, and consequently their position in the HR diagram is not sensitive to the choice of $\alpha_\mathrm{PMS}$. Moreover, we can not achieve a satisfactory agreement between our model and the primary star, as already pointed out by \citet{gennaro11}. Similarly to ASAS~J052821+0338.5, RXJ 0529.4+0041 A has two stars near the `heel'. As shown in Fig. \ref{fig:hr_bin}, the three different $\alpha_\mathrm{PMS}$ are all compatible with both stellar components within the observational uncertainties, which are quite large. 

V1174 Ori is much more problematic. As discussed in \citet{gennaro11}, the two stars show a peculiar position in the HR diagram. None of the present models (or other models widely adopted in the literature) can reproduce the correct position of the secondary by adopting the measured mass and chemical composition. A possible explanation of such a peculiar position in the HR diagram can be the presence of a large systematic uncertainty introduced by the adopted spectral type-effective temperature scale \citep[see e.g.,][]{luhman97,stassun04, hillenbrand04, gennaro11}. To be close to our coolest model ($\alpha_\mathrm{PMS}=1.0$), an increase of about 300 K in the secondary effective temperature would be required, which would correspond to a primary effective temperature increment of about 400 K. However, it seems unlikely that such a large shift could be caused uniquely by the adoption of a inadequate spectral type-effective temperature scale.

Figure \ref{fig:lit_bin} shows the comparisons between theoretical and observed lithium surface abundances. The evolutionary track of surface lithium abundance is shown. In the case of EK Cep we do not show the primary because the lithium abundance is not currently available. The figure shows the tracks computed with the low-convection efficiency, i.e. $\alpha_\mathrm{PMS}=1.0$ and the models with $\alpha_\mathrm{PMS}=\alpha_\mathrm{MS}=1.68$.

The lithium predictions of our models computed with both $\alpha_\mathrm{PMS}=1.0$ and $\alpha_\mathrm{PMS}=\alpha_\mathrm{MS}=1.68$ are in good agreement with data for the primary components (1.0 M$_\odot\la$ M$_1$ $\la 1.4$ M$_\odot$), within the uncertainties, since lithium depletion is almost negligible for such masses. Therefore, the primary components belonging to our sample do not allow further constraints on the $\alpha_\mathrm{PMS}$ value. In contrast, the impact of $\alpha_\mathrm{PMS}$ gets stronger and stronger as the mass decreases below about 1 M$_\odot$. The secondary of EK Cep and RXJ 0529.4+0041 A might thus give useful constraints\footnote{The case of V1174 Ori is peculiar, and if the problem resides in the effective temperature determinations, then lithium abundances could also be affected by uncertainties greater than the quoted ones.}. Unfortunately, among the selected systems, the lithium data for the secondary components are quite uncertain, with errors as large as 0.5 dex (see Table \ref{tab:binarie}), avoiding a robust comparison. The model with a low-convection efficiency is fully compatible with the data for the secondary component of EK Cep, although we cannot exclude $\alpha_\mathrm{PMS} =1.68$, while in the case of RXJ 0529.4+0041 A, nothing can be concluded owing to extremely uncertain lithium abundance determination, which is only a lower limit \citep[see also the discussion in,][]{alcala00,covino01,dantona03}. From this analysis it is evident that an effort to improve the measurements of the main parameters of binary systems, lithium abundance included, is required, in order to better constrain the super-adiabatic efficiency of theoretical models. Moreover, we emphasize that EBs are extremely useful tools for testing the validity of stellar models, and consequently, precise data about such systems would be required.

\section{Conclusions}

We have discussed in detail the uncertainties on theoretical models by evaluating the effect of the errors affecting the initial chemical composition and the up-to-date physical inputs (i.e. opacity, reaction rates, EOS), for several ages, masses, and chemical compositions. From these computations, we obtained a  quantitative estimation of the error bar associated to the $^7$Li surface abundance predictions. 
The comparison between theory and observations was conducted on five young clusters, namely Ic 2602, $\alpha$ Per, Blanco1, Pleiades, and Ngc 2516, and on four pre-MS EBs.

Our results confirm the disagreement between present standard models and $^7$Li surface abundance in young stars. Motivated by the high sensitivity of $^7$Li surface depletion to the convection efficiency and by the possibility of a dependency of the mixing length parameter on the evolutionary phase, gravity and $T_\mathrm{eff}$, we therefore explored the effect of a different mixing length parameter during the pre-MS and the MS evolution. We found that, in this case, very good agreement can be achieved for all clusters of the sample by adopting $\alpha_\mathrm{PMS} = 1.0$, which is considerably lower than the MS one (i.e. $\alpha_\mathrm{MS} = 1.68$ - 1.9), obtained by comparing theoretical isochrones with the observed colour-magnitude diagrams.

We also checked the validity of such low-convection efficiency models against four pre-MS detached double-lined eclipsing binaries, namely ASAS J052821+0338.5, EK Cep, RXJ~0529.4+0041 A, and V1174 Ori. We found that, in the HR diagram, pre-MS tracks with $\alpha = 1.0$ seem to agree well with the data for at least two of the three systems that have a star close to the Hayashi track (ASAS J052821+0338.5 and RXJ~0529.4+0041 A). However, the models are compatible with $^7$Li data both adopting the low- and high-convection efficiency, as a consequence of the large observational uncertainty still present on $^7$Li abundance determinations. 

Our results point out the necessity of low-convection efficiency during the pre-MS phase in standard models. However, further analysis is required to clarify whether the mixing length parameter actually changes from low to high values evolving from the pre-MS to the MS phase, or if some physical mechanism that acts in a way to partially inhibit convection in the envelope is lacking \citep[i.e. magnetic field, see e.g., ][]{ventura98}.

\begin{acknowledgements}
It is a great pleasure to thank Matteo Dell'Omodarme for his invaluable help in computer programming and Paola Sestito for useful and pleasant discussions. This research made use of the WEBDA database, operated at the Institute for Astronomy of the University of Vienna. This research has been partially supported by the PRIN-INAF 2011 (\emph{Tracing the formation and evolution of the galactic halo with VST}, P.I. Marcella Marconi).
\end{acknowledgements}

\bibliographystyle{aa} 
\bibliography{bibliografia_litio}

\begin{thebibliography}{99}
\expandafter\ifx\csname natexlab\endcsname\relax\def\natexlab#1{#1}\fi

\bibitem[{{Alcal{\'a}} {et~al.}(2000){Alcal{\'a}}, {Covino}, {Torres},
  {Sterzik}, {Pfeiffer}, \& {Neuh{\"a}user}}]{alcala00}
{Alcal{\'a}}, J.~M., {Covino}, E., {Torres}, G., {et~al.} 2000, \aap, 353, 186

\bibitem[{{Angulo} {et~al.}(1999){Angulo}, {Arnould}, {Rayet}, {Descouvemont},
  {Baye}, {Leclercq-Willain}, {Coc}, {Barhoumi}, {Aguer}, {Rolfs}, {Kunz},
  {Hammer}, {Mayer}, {Paradellis}, {Kossionides}, {Chronidou}, {Spyrou},
  {Degl'Innocenti}, {Fiorentini}, {Ricci}, {Zavatarelli}, {Providencia},
  {Wolters}, {Soares}, {Grama}, {Rahighi}, {Shotter}, \& {Lamehi
  Rachti}}]{nacre}
{Angulo}, C., {Arnould}, M., {Rayet}, M., {et~al.} 1999, \nphysa, 656, 3

\bibitem[{{Asplund} {et~al.}(2005){Asplund}, {Grevesse}, \&
  {Sauval}}]{asplund05}
{Asplund}, M., {Grevesse}, N., \& {Sauval}, A.~J. 2005, in Astronomical Society
  of the Pacific Conference Series, Vol. 336, Cosmic Abundances as Records of
  Stellar Evolution and Nucleosynthesis, ed. T.~G. {Barnes}, III \& F.~N.
  {Bash}, 25

\bibitem[{{Asplund} {et~al.}(2009){Asplund}, {Grevesse}, {Sauval}, \&
  {Scott}}]{asplund09}
{Asplund}, M., {Grevesse}, N., {Sauval}, A.~J., \& {Scott}, P. 2009, \araa, 47,
  481

\bibitem[{{Badnell} {et~al.}(2005){Badnell}, {Bautista}, {Butler}, {Delahaye},
  {Mendoza}, {Palmeri}, {Zeippen}, \& {Seaton}}]{badnell05}
{Badnell}, N.~R., {Bautista}, M.~A., {Butler}, K., {et~al.} 2005, \mnras, 360,
  458

\bibitem[{{Bahcall} \& {Serenelli}(2005)}]{bahcall05}
{Bahcall}, J.~N. \& {Serenelli}, A.~M. 2005, \apj, 626, 530

\bibitem[{{Bahcall} {et~al.}(2004){Bahcall}, {Serenelli}, \&
  {Pinsonneault}}]{bahcall04}
{Bahcall}, J.~N., {Serenelli}, A.~M., \& {Pinsonneault}, M. 2004, \apj, 614,
  464

\bibitem[{{Balachandran} {et~al.}(2011){Balachandran}, {Mallik}, \&
  {Lambert}}]{balachandran11}
{Balachandran}, S.~C., {Mallik}, S.~V., \& {Lambert}, D.~L. 2011, \mnras, 410,
  2526

\bibitem[{{Baraffe} \& {Chabrier}(2010)}]{baraffe10}
{Baraffe}, I. \& {Chabrier}, G. 2010, \aap, 521, A44

\bibitem[{{Boesgaard} \& {Tripicco}(1986)}]{boesgaard86}
{Boesgaard}, A.~M. \& {Tripicco}, M.~J. 1986, \apjl, 302, L49

\bibitem[{{B{\"o}hm-Vitense}(1958)}]{bohm58}
{B{\"o}hm-Vitense}, E. 1958, Zeitschrift fur Astrophysik, 46, 108

\bibitem[{{Bonaca} {et~al.}(2012){Bonaca}, {Tanner}, {Basu}, {Chaplin},
  {Metcalfe}, {Monteiro}, {Ballot}, {Bedding}, {Bonanno}, {Broomhall},
  {Bruntt}, {Campante}, {Christensen-Dalsgaard}, {Corsaro}, {Elsworth},
  {Garc{\'{\i}}a}, {Hekker}, {Karoff}, {Kjeldsen}, {Mathur}, {R{\'e}gulo},
  {Roxburgh}, {Stello}, {Trampedach}, {Barclay}, {Burke}, \&
  {Caldwell}}]{bonaca12}
{Bonaca}, A., {Tanner}, J.~D., {Basu}, S., {et~al.} 2012, ArXiv e-prints

\bibitem[{{Brocato} {et~al.}(2003){Brocato}, {Castellani}, {Di Carlo},
  {Raimondo}, \& {Walker}}]{brocato03}
{Brocato}, E., {Castellani}, V., {Di Carlo}, E., {Raimondo}, G., \& {Walker},
  A.~R. 2003, \aj, 125, 3111

\bibitem[{{Caffau} {et~al.}(2010){Caffau}, {Ludwig}, {Bonifacio}, {Faraggiana},
  {Steffen}, {Freytag}, {Kamp}, \& {Ayres}}]{caffau10}
{Caffau}, E., {Ludwig}, H.-G., {Bonifacio}, P., {et~al.} 2010, \aap, 514, A92

\bibitem[{{Casagrande} {et~al.}(2007){Casagrande}, {Flynn}, {Portinari},
  {Girardi}, \& {Jimenez}}]{casagrande07}
{Casagrande}, L., {Flynn}, C., {Portinari}, L., {Girardi}, L., \& {Jimenez}, R.
  2007, \mnras, 382, 1516

\bibitem[{{Chaboyer} {et~al.}(1995){Chaboyer}, {Demarque}, \&
  {Pinsonneault}}]{chaboyer95}
{Chaboyer}, B., {Demarque}, P., \& {Pinsonneault}, M.~H. 1995, \apj, 441, 876

\bibitem[{{Chabrier} \& {Baraffe}(1997)}]{chabrier97}
{Chabrier}, G. \& {Baraffe}, I. 1997, \aap, 327, 1039

\bibitem[{{Charbonnel} {et~al.}(2000){Charbonnel}, {Deliyannis}, \&
  {Pinsonneault}}]{charbonnel00}
{Charbonnel}, C., {Deliyannis}, C.~P., \& {Pinsonneault}, M. 2000, in IAU
  Symposium, Vol. 198, The Light Elements and their Evolution, ed. {L.~da
  Silva, R.~de Medeiros, \& M.~Spite}, 87

\bibitem[{{Charbonnel} \& {Talon}(2005)}]{charbonnel05}
{Charbonnel}, C. \& {Talon}, S. 2005, Science, 309, 2189

\bibitem[{{Chieffi} {et~al.}(1995){Chieffi}, {Straniero}, \&
  {Salaris}}]{chieffi95}
{Chieffi}, A., {Straniero}, O., \& {Salaris}, M. 1995, \apjl, 445, L39

\bibitem[{{Claret}(2007)}]{claret07}
{Claret}, A. 2007, \aap, 475, 1019

\bibitem[{{Clarke} {et~al.}(2004){Clarke}, {MacDonald}, \& {Owens}}]{clarke04}
{Clarke}, D., {MacDonald}, E.~C., \& {Owens}, S. 2004, \aap, 415, 677

\bibitem[{{Covino} {et~al.}(2000){Covino}, {Catalano}, {Frasca}, {Marilli},
  {Fern{\'a}ndez}, {Alcal{\'a}}, {Melo}, {Paladino}, {Sterzik}, \&
  {Stelzer}}]{covino00}
{Covino}, E., {Catalano}, S., {Frasca}, A., {et~al.} 2000, \aap, 361, L49

\bibitem[{{Covino} {et~al.}(2004){Covino}, {Frasca}, {Alcal{\'a}}, {Paladino},
  \& {Sterzik}}]{covino04}
{Covino}, E., {Frasca}, A., {Alcal{\'a}}, J.~M., {Paladino}, R., \& {Sterzik},
  M.~F. 2004, \aap, 427, 637

\bibitem[{{Covino} {et~al.}(2001){Covino}, {Melo}, {Alcal{\'a}}, {Torres},
  {Fern{\'a}ndez}, {Frasca}, \& {Paladino}}]{covino01}
{Covino}, E., {Melo}, C., {Alcal{\'a}}, J.~M., {et~al.} 2001, \aap, 375, 130

\bibitem[{{Cox} \& {Giuli}(1968)}]{cox}
{Cox}, J.~P. \& {Giuli}, R.~T. 1968, {Principles of stellar structure} (New
  York, Gordon and Breach [1968])

\bibitem[{{Cyburt} {et~al.}(2004){Cyburt}, {Fields}, \& {Olive}}]{cyburt04}
{Cyburt}, R.~H., {Fields}, B.~D., \& {Olive}, K.~A. 2004, \prd, 69, 123519

\bibitem[{{D'Antona}(1993)}]{dantona93}
{D'Antona}, F. 1993, in Astronomical Society of the Pacific Conference Series,
  Vol.~40, IAU Colloq. 137: Inside the Stars, ed. {W.~W.~Weiss \& A.~Baglin},
  395--409

\bibitem[{{D'Antona} \& {Mazzitelli}(1984)}]{dantona84}
{D'Antona}, F. \& {Mazzitelli}, I. 1984, \aap, 138, 431

\bibitem[{{D'Antona} \& {Mazzitelli}(1997)}]{dantona97}
{D'Antona}, F. \& {Mazzitelli}, I. 1997, \memsai, 68, 807

\bibitem[{{D'Antona} \& {Montalb{\'a}n}(2003)}]{dantona03}
{D'Antona}, F. \& {Montalb{\'a}n}, J. 2003, \aap, 412, 213

\bibitem[{{D'Antona} {et~al.}(2000){D'Antona}, {Ventura}, \&
  {Mazzitelli}}]{dantona00}
{D'Antona}, F., {Ventura}, P., \& {Mazzitelli}, I. 2000, \apjl, 543, L77

\bibitem[{{Degl'Innocenti} {et~al.}(2008){Degl'Innocenti}, {Prada Moroni},
  {Marconi}, \& {Ruoppo}}]{deglinnocenti08}
{Degl'Innocenti}, S., {Prada Moroni}, P.~G., {Marconi}, M., \& {Ruoppo}, A.
  2008, \apss, 316, 25

\bibitem[{{Deliyannis} {et~al.}(2000){Deliyannis}, {Pinsonneault}, \&
  {Charbonnel}}]{deliyannis00}
{Deliyannis}, C.~P., {Pinsonneault}, M.~H., \& {Charbonnel}, C. 2000, in IAU
  Symposium, Vol. 198, The Light Elements and their Evolution, ed. L.~{da
  Silva}, R.~{de Medeiros}, \& M.~{Spite}, 61

\bibitem[{{di Criscienzo} {et~al.}(2010){di Criscienzo}, {Ventura}, \&
  {D'Antona}}]{dicriscienzo10}
{di Criscienzo}, M., {Ventura}, P., \& {D'Antona}, F. 2010, \apss, 328, 167

\bibitem[{{D'Orazi} \& {Randich}(2009)}]{dorazi09}
{D'Orazi}, V. \& {Randich}, S. 2009, \aap, 501, 553

\bibitem[{{D'Orazi} {et~al.}(2009){D'Orazi}, {Randich}, {Flaccomio}, {Palla},
  {Sacco}, \& {Pallavicini}}]{dorazi09b}
{D'Orazi}, V., {Randich}, S., {Flaccomio}, E., {et~al.} 2009, \aap, 501, 973

\bibitem[{{Dorman} {et~al.}(1989){Dorman}, {Nelson}, \& {Chau}}]{dorman89}
{Dorman}, B., {Nelson}, L.~A., \& {Chau}, W.~Y. 1989, \apj, 342, 1003

\bibitem[{{Ferguson} {et~al.}(2005){Ferguson}, {Alexander}, {Allard}, {Barman},
  {Bodnarik}, {Hauschildt}, {Heffner-Wong}, \& {Tamanai}}]{ferguson05}
{Ferguson}, J.~W., {Alexander}, D.~R., {Allard}, F., {et~al.} 2005, \apj, 623,
  585

\bibitem[{{Ferraro} {et~al.}(2006){Ferraro}, {Valenti}, {Straniero}, \&
  {Origlia}}]{ferraro06}
{Ferraro}, F.~R., {Valenti}, E., {Straniero}, O., \& {Origlia}, L. 2006, \apj,
  642, 225

\bibitem[{{Ford} {et~al.}(2005){Ford}, {Jeffries}, \& {Smalley}}]{ford05}
{Ford}, A., {Jeffries}, R.~D., \& {Smalley}, B. 2005, \mnras, 364, 272

\bibitem[{{Geiss} \& {Gloeckler}(1998)}]{geiss98}
{Geiss}, J. \& {Gloeckler}, G. 1998, Space Science Reviews, 84, 239

\bibitem[{{Gennaro} {et~al.}(2010){Gennaro}, {Prada Moroni}, \&
  {Degl'Innocenti}}]{gennaro10}
{Gennaro}, M., {Prada Moroni}, P.~G., \& {Degl'Innocenti}, S. 2010, \aap, 518,
  A13

\bibitem[{{Gennaro} {et~al.}(2011){Gennaro}, {Prada Moroni}, \&
  {Tognelli}}]{gennaro11}
{Gennaro}, M., {Prada Moroni}, P.~G., \& {Tognelli}, E. 2011, ArXiv e-prints

\bibitem[{{Grevesse} \& {Noels}(1993)}]{grevesse93}
{Grevesse}, N. \& {Noels}, A. 1993, in Origin and Evolution of the Elements,
  ed. {N.~Prantzos, E.~Vangioni-Flam, \& M.~Casse}, 15--25

\bibitem[{{Grevesse} \& {Sauval}(1998)}]{grevesse98}
{Grevesse}, N. \& {Sauval}, A.~J. 1998, \ssr, 85, 161

\bibitem[{{Hillenbrand} \& {White}(2004)}]{hillenbrand04}
{Hillenbrand}, L.~A. \& {White}, R.~J. 2004, \apj, 604, 741

\bibitem[{{Iglesias} \& {Rogers}(1996)}]{iglesias96}
{Iglesias}, C.~A. \& {Rogers}, F.~J. 1996, \apj, 464, 943

\bibitem[{{Jeffries}(2000)}]{jeffries00}
{Jeffries}, R.~D. 2000, in Astronomical Society of the Pacific Conference
  Series, Vol. 198, Stellar Clusters and Associations: Convection, Rotation,
  and Dynamos, ed. R.~{Pallavicini}, G.~{Micela}, \& S.~{Sciortino}, 245

\bibitem[{{Jeffries}(2006)}]{jeffries06}
{Jeffries}, R.~D. 2006, {Pre-Main-Sequence Lithium Depletion} (Chemical
  Abundances and Mixing in Stars in the Milky Way and its Satellites, ESO
  ASTROPHYSICS SYMPOSIA.~ISBN 978-3-540-34135-2.~Springer-Verlag, 2006,
  p.~163), 163

\bibitem[{{King} {et~al.}(2000){King}, {Krishnamurthi}, \&
  {Pinsonneault}}]{king00}
{King}, J.~R., {Krishnamurthi}, A., \& {Pinsonneault}, M.~H. 2000, \aj, 119,
  859

\bibitem[{{King} {et~al.}(2010){King}, {Schuler}, {Hobbs}, \&
  {Pinsonneault}}]{king10}
{King}, J.~R., {Schuler}, S.~C., {Hobbs}, L.~M., \& {Pinsonneault}, M.~H. 2010,
  \apj, 710, 1610

\bibitem[{{Landin} {et~al.}(2006){Landin}, {Ventura}, {D'Antona}, {Mendes}, \&
  {Vaz}}]{landin06}
{Landin}, N.~R., {Ventura}, P., {D'Antona}, F., {Mendes}, L.~T.~S., \& {Vaz},
  L.~P.~R. 2006, \aap, 456, 269

\bibitem[{{Lattuada} {et~al.}(2001){Lattuada}, {Pizzone}, {Typel}, {Figuera},
  {Miljani{\'c}}, {Musumarra}, {Pellegriti}, {Rolfs}, {Spitaleri}, \&
  {Wolter}}]{lattuada01}
{Lattuada}, M., {Pizzone}, R.~G., {Typel}, S., {et~al.} 2001, \apj, 562, 1076

\bibitem[{{Linsky} {et~al.}(2006){Linsky}, {Draine}, {Moos}, {Jenkins}, {Wood},
  {Oliveira}, {Blair}, {Friedman}, {Gry}, {Knauth}, {Kruk}, {Lacour}, {Lehner},
  {Redfield}, {Shull}, {Sonneborn}, \& {Williger}}]{linsky06}
{Linsky}, J.~L., {Draine}, B.~T., {Moos}, H.~W., {et~al.} 2006, \apj, 647, 1106

\bibitem[{{Lodders} {et~al.}(2009){Lodders}, {Plame}, \& {Gail}}]{lodders09}
{Lodders}, K., {Plame}, H., \& {Gail}, H. 2009, in Landolt-B{\"o}rnstein -
  Group VI Astronomy and Astrophysics Numerical Data and Functional
  Relationships in Science and Technology Volume 4B: Solar System. Edited by
  J.E. Tr{\"u}mper, 2009, 4.4., 44

\bibitem[{{Ludwig} {et~al.}(1999){Ludwig}, {Freytag}, \& {Steffen}}]{ludwig99}
{Ludwig}, H.-G., {Freytag}, B., \& {Steffen}, M. 1999, \aap, 346, 111

\bibitem[{{Luhman} {et~al.}(1997){Luhman}, {Liebert}, \& {Rieke}}]{luhman97}
{Luhman}, K.~L., {Liebert}, J., \& {Rieke}, G.~H. 1997, \apjl, 489, L165

\bibitem[{{Martin} \& {Rebolo}(1993)}]{martin93}
{Martin}, E.~L. \& {Rebolo}, R. 1993, \aap, 274, 274

\bibitem[{{Mathieu} {et~al.}(2007){Mathieu}, {Baraffe}, {Simon}, {Stassun}, \&
  {White}}]{mathieu07}
{Mathieu}, R.~D., {Baraffe}, I., {Simon}, M., {Stassun}, K.~G., \& {White}, R.
  2007, Protostars and Planets V, 411

\bibitem[{{Mazzitelli}(1989)}]{mazzitelli89}
{Mazzitelli}, I. 1989, in European Southern Observatory Conference and Workshop
  Proceedings, Vol.~33, European Southern Observatory Conference and Workshop
  Proceedings, ed. {B.~Reipurth}, 433--445

\bibitem[{{Mendes} {et~al.}(1999){Mendes}, {D'Antona}, \&
  {Mazzitelli}}]{mendes99}
{Mendes}, L.~T.~S., {D'Antona}, F., \& {Mazzitelli}, I. 1999, \aap, 341, 174

\bibitem[{{Montalb{\'a}n} \& {D'Antona}(2006)}]{montalban06}
{Montalb{\'a}n}, J. \& {D'Antona}, F. 2006, \mnras, 370, 1823

\bibitem[{{Morel} {et~al.}(2000){Morel}, {Morel}, {Provost}, \&
  {Berthomieu}}]{morel00}
{Morel}, P., {Morel}, C., {Provost}, J., \& {Berthomieu}, G. 2000, \aap, 354,
  636

\bibitem[{{Neece}(1984)}]{neece84}
{Neece}, G.~D. 1984, \apj, 277, 738

\bibitem[{{Neuforge-Verheecke} {et~al.}(2001){Neuforge-Verheecke}, {Guzik},
  {Keady}, {Magee}, {Bradley}, \& {Noels}}]{neuforge01}
{Neuforge-Verheecke}, C., {Guzik}, J.~A., {Keady}, J.~J., {et~al.} 2001, \apj,
  561, 450

\bibitem[{{Pace} {et~al.}(2012){Pace}, {Castro}, {Mel{\'e}ndez}, {Th{\'e}ado},
  \& {do Nascimento}}]{pace12}
{Pace}, G., {Castro}, M., {Mel{\'e}ndez}, J., {Th{\'e}ado}, S., \& {do
  Nascimento}, Jr., J.-D. 2012, \aap, 541, A150

\bibitem[{{Piau} {et~al.}(2011){Piau}, {Kervella}, {Dib}, \&
  {Hauschildt}}]{piau11}
{Piau}, L., {Kervella}, P., {Dib}, S., \& {Hauschildt}, P. 2011, \aap, 526,
  A100

\bibitem[{{Piau} \& {Turck-Chi{\`e}ze}(2002)}]{piau02}
{Piau}, L. \& {Turck-Chi{\`e}ze}, S. 2002, \apj, 566, 419

\bibitem[{{Pinsonneault}(1994)}]{pinsonneault94}
{Pinsonneault}, M.~H. 1994, in Astronomical Society of the Pacific Conference
  Series, Vol.~64, Cool Stars, Stellar Systems, and the Sun, ed.
  {J.-P.~Caillault}, 254

\bibitem[{{Pinsonneault} {et~al.}(2000){Pinsonneault}, {Charbonnel}, \&
  {Deliyannis}}]{pinsonneault00}
{Pinsonneault}, M.~H., {Charbonnel}, C., \& {Deliyannis}, C.~P. 2000, in IAU
  Symposium, Vol. 198, The Light Elements and their Evolution, ed. {L.~da
  Silva, R.~de Medeiros, \& M.~Spite}, 74

\bibitem[{{Pinsonneault} {et~al.}(1990){Pinsonneault}, {Kawaler}, \&
  {Demarque}}]{pinsonneault90}
{Pinsonneault}, M.~H., {Kawaler}, S.~D., \& {Demarque}, P. 1990, \apjs, 74, 501

\bibitem[{{Pols} {et~al.}(1995){Pols}, {Tout}, {Eggleton}, \& {Han}}]{pols95}
{Pols}, O.~R., {Tout}, C.~A., {Eggleton}, P.~P., \& {Han}, Z. 1995, \mnras,
  274, 964

\bibitem[{{Popper}(1987)}]{popper87}
{Popper}, D.~M. 1987, \apjl, 313, L81

\bibitem[{{Richer} \& {Michaud}(1993)}]{richer93}
{Richer}, J. \& {Michaud}, G. 1993, \apj, 416, 312

\bibitem[{{Rogers} \& {Iglesias}(1992)}]{rogers92}
{Rogers}, F.~J. \& {Iglesias}, C.~A. 1992, \apjs, 79, 507

\bibitem[{{Seaton} {et~al.}(1994){Seaton}, {Yan}, {Mihalas}, \&
  {Pradhan}}]{seaton94}
{Seaton}, M.~J., {Yan}, Y., {Mihalas}, D., \& {Pradhan}, A.~K. 1994, \mnras,
  266, 805

\bibitem[{{Sestito} {et~al.}(2006){Sestito}, {Degl'Innocenti}, {Prada Moroni},
  \& {Randich}}]{sestito06}
{Sestito}, P., {Degl'Innocenti}, S., {Prada Moroni}, P.~G., \& {Randich}, S.
  2006, \aap, 454, 311

\bibitem[{{Sestito} \& {Randich}(2005)}]{sestito05}
{Sestito}, P. \& {Randich}, S. 2005, \aap, 442, 615

\bibitem[{{Siess} {et~al.}(1999){Siess}, {Forestini}, \& {Bertout}}]{siess99}
{Siess}, L., {Forestini}, M., \& {Bertout}, C. 1999, \aap, 342, 480

\bibitem[{{Soderblom} {et~al.}(2009){Soderblom}, {Laskar}, {Valenti},
  {Stauffer}, \& {Rebull}}]{soderblom09}
{Soderblom}, D.~R., {Laskar}, T., {Valenti}, J.~A., {Stauffer}, J.~R., \&
  {Rebull}, L.~M. 2009, \aj, 138, 1292

\bibitem[{{Stassun} {et~al.}(2004){Stassun}, {Mathieu}, {Vaz}, {Stroud}, \&
  {Vrba}}]{stassun04}
{Stassun}, K.~G., {Mathieu}, R.~D., {Vaz}, L.~P.~R., {Stroud}, N., \& {Vrba},
  F.~J. 2004, \apjs, 151, 357

\bibitem[{{Steigman} {et~al.}(2007){Steigman}, {Romano}, \&
  {Tosi}}]{steigman07}
{Steigman}, G., {Romano}, D., \& {Tosi}, M. 2007, \mnras, 378, 576

\bibitem[{{Stempels} {et~al.}(2008){Stempels}, {Hebb}, {Stassun}, {Holtzman},
  {Dunstone}, {Glowienka}, \& {Frandsen}}]{stempels08}
{Stempels}, H.~C., {Hebb}, L., {Stassun}, K.~G., {et~al.} 2008, \aap, 481, 747

\bibitem[{{Sung} {et~al.}(2002){Sung}, {Bessell}, {Lee}, \& {Lee}}]{sung02}
{Sung}, H., {Bessell}, M.~S., {Lee}, B.-W., \& {Lee}, S.-G. 2002, \aj, 123, 290

\bibitem[{{Swenson} {et~al.}(1994){Swenson}, {Faulkner}, {Rogers}, \&
  {Iglesias}}]{swenson94}
{Swenson}, F.~J., {Faulkner}, J., {Rogers}, F.~J., \& {Iglesias}, C.~A. 1994,
  \apj, 425, 286

\bibitem[{{Talon}(2008)}]{talon08}
{Talon}, S. 2008, in EAS Publications Series, Vol.~32, EAS Publications Series,
  ed. C.~{Charbonnel} \& J.-P. {Zahn}, 81--130

\bibitem[{{Talon} \& {Charbonnel}(1998)}]{talon98}
{Talon}, S. \& {Charbonnel}, C. 1998, \aap, 335, 959

\bibitem[{{Talon} \& {Charbonnel}(2010)}]{talon10}
{Talon}, S. \& {Charbonnel}, C. 2010, in IAU Symposium, Vol. 268, IAU
  Symposium, ed. {C.~Charbonnel, M.~Tosi, F.~Primas, \& C.~Chiappini}, 365--374

\bibitem[{{Terndrup} {et~al.}(2002){Terndrup}, {Pinsonneault}, {Jeffries},
  {Ford}, {Stauffer}, \& {Sills}}]{terndrup02}
{Terndrup}, D.~M., {Pinsonneault}, M., {Jeffries}, R.~D., {et~al.} 2002, \apj,
  576, 950

\bibitem[{{Tognelli} {et~al.}(2011){Tognelli}, {Prada Moroni}, \&
  {Degl'Innocenti}}]{tognelli11}
{Tognelli}, E., {Prada Moroni}, P.~G., \& {Degl'Innocenti}, S. 2011, \aap, 533,
  A109

\bibitem[{{Trampedach}(2007)}]{trampedach07}
{Trampedach}, R. 2007, in American Institute of Physics Conference Series, Vol.
  948, Unsolved Problems in Stellar Physics: A Conference in Honor of Douglas
  Gough, ed. {R.~J.~Stancliffe, G.~Houdek, R.~G.~Martin, \& C.~A.~Tout},
  141--148

\bibitem[{{Trampedach} {et~al.}(2006){Trampedach}, {D{\"a}ppen}, \&
  {Baturin}}]{trampedach06}
{Trampedach}, R., {D{\"a}ppen}, W., \& {Baturin}, V.~A. 2006, \apj, 646, 560

\bibitem[{{Umezu} \& {Saio}(2000)}]{umezu00}
{Umezu}, M. \& {Saio}, H. 2000, \mnras, 316, 307

\bibitem[{{Valle} {et~al.}(2012){Valle}, {Dell'Omodarme}, {Prada Moroni}, \&
  {Degl'Innocenti}}]{valle12}
{Valle}, G., {Dell'Omodarme}, M., {Prada Moroni}, P.~G., \& {Degl'Innocenti},
  S. 2012, \aap, Submitted

\bibitem[{{Ventura} {et~al.}(1998){Ventura}, {Zeppieri}, {Mazzitelli}, \&
  {D'Antona}}]{ventura98}
{Ventura}, P., {Zeppieri}, A., {Mazzitelli}, I., \& {D'Antona}, F. 1998, \aap,
  331, 1011

\bibitem[{{Vick} {et~al.}(2010){Vick}, {Michaud}, {Richer}, \&
  {Richard}}]{vick10}
{Vick}, M., {Michaud}, G., {Richer}, J., \& {Richard}, O. 2010, \aap, 521, A62

\bibitem[{{Xiong} \& {Deng}(2006)}]{xiong06}
{Xiong}, D.-R. \& {Deng}, L.-C. 2006, \caa, 30, 24

\bibitem[{{Y{\i}ld{\i}z}(2007)}]{yildiz07}
{Y{\i}ld{\i}z}, M. 2007, \mnras, 374, 1264

\end{thebibliography}

\end{document}